\documentclass{elsarticle}
\usepackage{hyperref}
\usepackage[utf8]{inputenc}
\usepackage{wrapfig}
\usepackage{amsmath}
\usepackage{amsthm}
\usepackage{amssymb}
\usepackage[tmargin=1in,bmargin=1in]{geometry}
\usepackage{caption}
\usepackage{subcaption}
\usepackage{graphicx}
\usepackage{latexsym}
\usepackage[boxed]{algorithm}
\usepackage{algorithm}
\usepackage{algpseudocode}
\usepackage{multirow}
\usepackage{rotating}
\usepackage{color}
\usepackage{url}
\usepackage{bigstrut}
\usepackage[utf8]{inputenc}
\usepackage[overload]{empheq}
\usepackage{cases}  
\usepackage{ulem}
\usepackage{lineno}
\usepackage{longtable}
\usepackage[dvipsnames]{xcolor}

\hypersetup{
    colorlinks=true,              
    linkcolor=blue,               
    citecolor=blue,               
    filecolor=black,              
    urlcolor=red,               
    bookmarks=false,
    pdffitwindow=true,
    pdfpagelayout=SinglePage
}

\newcommand{\vect}[1]{\boldsymbol{#1}}

\newcommand{\mat}[1]{\underline{\underline{\boldsymbol{#1}}}}

\newcommand{\hf}{{\frac12}}

\newcommand{\cell}[1]{\mathcal{V}_{#1}}
\newcommand{\diff}[1]{\, d#1}
\newcommand{\Oof}[1]{\mathcal{O}\left( #1 \right)}

\newcommand{\ddn}[1]{\partial_{\vect{n}} #1}
\newcommand{\ddna}[2]{\partial_{\vect{n}^{#1}} #2}
\newcommand{\jump}[1]{\left[ #1 \right]}

\newcommand{\leftbrace}[1]{\left\{\begin{aligned}#1\end{aligned}\right.\end{align*}}

\usepackage{tikz}

\newcommand{\tm}[1]{{#1}}
\newcommand{\abs}[1]{\left| #1 \right|}
\usepackage{algpseudocode}

\algdef{SE}[DOWHILE]{Do}{doWhile}{\algorithmicdo}[1]{\algorithmicwhile\ #1}

\newcommand{\set}[3]{\left\{ #1 \right\}_{#2}^{#3}}
\usepackage{tikz}
\usetikzlibrary{math}
\usetikzlibrary{calc}

\tikzset{%
  colorLiquid/.style={fill=white, opacity=0.25},
  colorSolid/.style={fill=yellow, opacity=0.25},
  colorFront/.style={draw=orange, thick},
  pointingLine/.style={draw=black, opacity=0.4, thin},
  fontSizeLarge/.style={scale=1.0},
  fontSizeNorma/.style={scale=0.9},
  fontSizeSmall/.style={scale=0.8},
  fontColorAuxA/.style={text=blue!50!cyan},
  fontColorAuxB/.style={text=green!50!black},
  fontColorAuxC/.style={text=red!75!black}
}

\newcommand{\reviewerOne}[1]{\textcolor{black}{#1}}
\newcommand{\reviewerTwo}[1]{\textcolor{black}{#1}}

\newcommand{\ours}[1]{\textcolor{black}{#1}}

\newcommand{\mylinelabel}[1]{}

\begin{document}


\title{Solving Elliptic Interface Problems with Jump Conditions on \mylinelabel{rev2:1}\reviewerTwo{Cartesian} Grids}

\cortext[cor]{Corresponding author: dbochkov@ucsb.edu}

\address[MECHE]{Department of Mechanical Engineering, University of California, Santa Barbara, CA 93106}
\address[CS]{Department of Computer Science, University of California, Santa Barbara, CA 93106}

\author[MECHE]{Daniil Bochkov}
\author[MECHE,CS]{Frederic Gibou}

\begin{abstract}
We present a simple numerical algorithm for solving elliptic equations where the diffusion coefficient, the source term, the solution and its flux are discontinuous across an irregular interface. The algorithm produces second-order accurate solutions and first-order accurate gradients in the $L^\infty$-norm on Cartesian grids. The condition number is bounded, regardless of the ratio of the diffusion constant and scales like that of the standard 5-point stencil approximation on a rectangular grid with no interface. Numerical examples are given in two and three spatial dimensions.
\end{abstract}

\begin{keyword}
  Poisson equation, Immersed interface, Level-Set Method
\end{keyword}

\maketitle

\section{Introduction}
It is crucial, for simulating important processes in the physical and life sciences, to find the numerical solution of elliptic equations with discontinuities in the diffusion coefficient, the source term, the solution and its flux. In the case of interfacial flows for example, jump conditions describe the discontinuity in stress that is balanced by forces that exist between phases \cite{Brennen:2005aa}. In the simulation of protein folding, it is the electrostatic potential that has a jump across the protein's Solvent-Excluded Surface \cite{Egan:2018aa, Mirzadeh;Theillard;Helgadottir;etal:12:An-Adaptive-Finite-D, Mirzadeh;Theillard;Gibou:11:A-second-order-discr, Xia:2014aa}. Other examples include solidification of multicomponent alloys \cite{Theillard;Gibou;Pollock:14:A-Sharp-Computationa, Kurz;Fisher:98:Fundamentals-of-Soli, Bochkov;Gibou:19:A-Sharp-Computationa} or any diffusion dominated processes with different materials properties. At the macroscale, changes across the surface can only be represented by sharp jumps, hence the need to numerically represent them as such. Failure to do so introduces errors that change the characteristics of the problem.

Numerical approximations to solve such problems have been proposed and fall into two categories, depending on whether the interface is represented explicitly or implicitly. For example, finite element discretizations approximate the space in which the solution is defined and rely on a mesh that explicitly describes the surface \cite{Babuska:70:The-finite-element-m}. It is straightforward to impose boundary conditions in that framework, which is ideally suited for cases where deformations are small. For large deformations, difficulties associated with the mesh generation process are severe. Consequently, in this case, implicit representations of the interface have proved to be a better choice; imposing jump conditions, however, is a difficulty task in that framework. One of the first \mylinelabel{rev2:2}\reviewerTwo{attempts} is the Immersed Interface Method (IIM), where the \mylinelabel{rev2:3}\reviewerTwo{jump} conditions are combined with Taylor expansions of the solution on each side of the interface in order to modify the stencils of grid points adjacent to the interface. The main difficulties are the need to evaluate high-order jump conditions and surface derivatives. Several authors have further developed numerical methods within the IIM framework, e.g. \cite{Chen;Strain:08:Piecewise-polynomial, Li:98:A-Fast-Iterative-Alg, Li;Ito:06:The-Immersed-Interfa, Wiegmann;Bube:00:The-explicit-jump-im, Berthelsen:04:A-decomposed-immerse, Adams;Chartier:04:New-geometric-immers, Adams;Chartier:05:A-comparison-of-alge, Adams;Li:02:The-immersed-interfa, Theillard;Gibou;Pollock:14:A-Sharp-Computationa}. Another approach is the Ghost Fluid Method (GFM) \cite{Fedkiw1999}, first developed to treat shocks and contact discontinuities in compressible flows. The idea is to define a ghost fluid in the regions across the discontinuities by adding the interface jump to the true fluid. This simple treatment avoids the large error incurred by differentiating discontinuous solutions, and thus gives an elegant framework to manage jump conditions. The idea of the GFM was used for solving the Poisson equation with jump conditions in \cite{Liu;Fedkiw;Kang:00:A-boundary-capturing}. In this case, the jump in the normal derivative of the solution is projected onto the Cartesian directions in order to use a dimension-by-dimension approach. The authors showed that the normal jump is accurately captured, while the tangential jump is smeared; this, in turn, leads to a lack of convergence in the flux. The Voronoi Interface Method \cite{Guittet;Lepilliez;Tanguy;etal:15:Solving-elliptic-pro} solved that problem by first constructing a local Voronoi mesh adjacent to the interface and by then considering a GFM treatment. In that case, the solution is second-order accurate in the $L^{\infty}$-norm with first-order accurate fluxes in the same norm. This method has been applied to electroporation problems \cite{Guittet;Poignard;Gibou:17:A-Voronoi-Interface-, Mistani:2019aa}, where the unknown is the electric potential at each grid points. While this method produces symmetric positive definite linear systems and only requires the right-hand side of the linear system to be modified, it requires the generation of a local Voronoi mesh \reviewerTwo{and interpolation of numerical solutions from such unstructured meshes back onto Cartesian grids}, which may add some challenges, especially in three spatial dimensions. The literature on solving elliptic problems with jump conditions is quite vast and we refer the interested reader to the review \cite{Gibou;Min;Fedkiw:13:High-Resolution-Shar} and to other approaches, such as cut-cell approaches \cite{Crockett;Colella;Graves:11:A-Cartesian-grid-emb, Oevermann;Scharfenberg;Klein:09:A-sharp-interface-fi}, discontinuous Galerkin and the eXtended Finite Element Method (XFEM) \cite{Lew;Buscaglia:08:A-discontinuous-Gale, Guyomarch;Lee;Jeon:09:A-discontinuous-Gale, Moes;Dolbow;Belytschko:99:A-finite-element-met, Daux;Moes;Dolbow;etal:00:Arbitrary-branched-a, Belytschko;Moes;Usui;etal:01:Arbitrary-discontinu, Moes;Cloirec;Cartraud;etal:03:A-computational-appr, Ji;Dolbow:04:On-strategies-for-en, Fries;Belytschko:06:The-intrinsic-XFEM:-, Groi;Reusken:07:An-extended-pressure, Bos;Gravemeier:09:Numerical-simulation}, the Virtual Node Method \cite{Molino;Bao;Fedkiw:04:A-Virtual-Node-Algor, Bao;Hong;Teran;etal:07:Fracturing-rigid-mat, Molino;Bao;Fedkiw:04:A-Virtual-Node-Algor, Sifakis;Der;Fedkiw:07:Arbitrary-cutting-of, Richardson;Hegemann;Sifakis;etal:11:An-XFEM-method-for-m, Jr.;Wang;Sifakis;etal:12:A-second-order-virtu} or other fictitious domain approaches \cite{Coco;Russo:12:Second-Order-Multigr, Cisternino;Weynans:12:A-parallel-second-or, Gallinato:2017aa}.

In this work, we propose a finite volume discretization for elliptic interface problems in a similar vein as in \cite{Ng;Min;Gibou:09:An-efficient-fluid--, Papac;Gibou;Ratsch:10:Efficient-symmetric-,Bochkov;Gibou:19:Solving-the-Poisson-} for the treatment of Neuman and Robin boundary conditions. To take into account the jump conditions, we adopt the ideas of relating the values of discontinuous functions using Taylor expansions in the normal direction and employing \ours{one-sided} local least-square interpolations. \mylinelabel{ours:newrefs}\ours{In these aspects, the present method is similar to the variant of the augmented immersed interface method used in \cite{Hu;Lai;Young:15:A-hybrid-immersed-bo, Xu;Shi;Hu;Etal:20:A-level-set-immersed}. The differences are, however, substantial: first, a finite volume approach is used instead of finite differences; second, the resulting linear system contains no augmented variable, which makes it straightforward to invert with ``black-box'' linear solvers (BiCGSTAB is used in this work; in \cite{Hu;Lai;Young:15:A-hybrid-immersed-bo, Xu;Shi;Hu;Etal:20:A-level-set-immersed} the system is solved iteratively using GMRES in conjunction with a fast Poisson solver); third, the method does not involve quadratic terms in Taylor expansions and local interpolations are linear (compared to cubic ones in \cite{Hu;Lai;Young:15:A-hybrid-immersed-bo, Xu;Shi;Hu;Etal:20:A-level-set-immersed}), which keeps the discretization stencil quite compact while still resulting in second-order accurate solutions (thanks to the finite volume approach). The small stencil size and the simplicity of inverting the resulting linear system make the method a good candidate for parallelization and application in the context of adaptive grids, which will be demonstrated in this work as well.}  We consider a level-set representation of the interface so that the method can be used in free boundary problems \cite{Osher;Sethian:88:Fronts-propagating-w, Osher;Fedkiw:03:Level-Set-Methods-an, Sethian:96:Level-set-methods, Gibou:2018aa}.

\section{Numerical Discretization}

\begin{figure}[h]
  \centering
  \begin{subfigure}[b]{0.32\textwidth}
    \centering
    \includegraphics[width=0.95\textwidth]{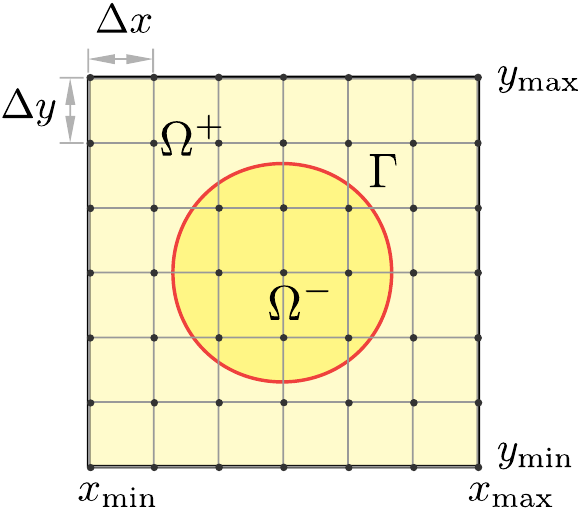}
    \caption{}
    \label{fig::domain_and_grid}
  \end{subfigure}
  \begin{subfigure}[b]{0.32\textwidth}
    \centering
    \includegraphics[width=0.75\textwidth]{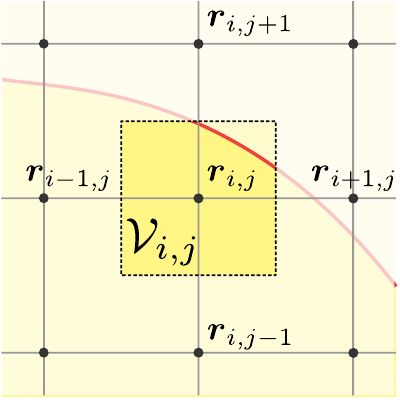}
    \caption{}
    \label{fig::finite_volume}
  \end{subfigure}
  \begin{subfigure}[b]{0.32\textwidth}
    \centering
    \includegraphics[width=0.75\textwidth]{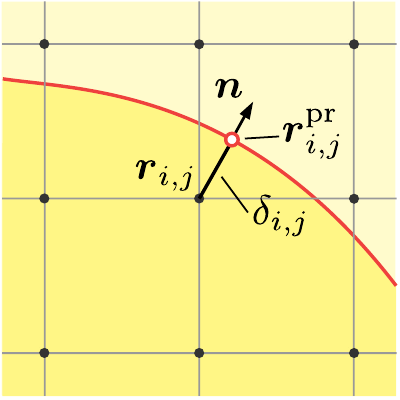}
    \caption{}
    \label{fig::projection}
  \end{subfigure}
  \caption{(a) Notation used in this paper. (b) Illustration of a finite volume associated with a grid point $(i, j)$. (c) Illustration of the projection of a grid point onto the interface $\Gamma$}
\end{figure}

Consider a rectangular domain $\Omega = [x_{\min}; x_{\max}]\times[y_{\min}; y_{\max}]$ with an immersed irregular interface $\Gamma$ that splits $\Omega$ into two sets $\Omega^-$ and $\Omega^+$ as illustrated in Fig. \ref{fig::domain_and_grid}. We seek a numerical solution $u=u(\vect{r})$, with $\vect{r} = (x, y)$, to the following problem:
\begin{linenomath*}
\begin{alignat}{2}
  \label{eq::poisson}
  k^\pm u^\pm - \nabla \cdot \left( \mu^\pm \nabla u^\pm \right) &= f^\pm, \quad &\text{in } &\Omega^\pm,
  \\
  \label{eq::jump_value}
  \jump{u} &= \alpha, \quad &\text{on } &\Gamma,
  \\
  \label{eq::jump_flux}
  \jump{\mu \ddn{u}} &= \beta, &\quad \text{on } &\Gamma,
\end{alignat}
\end{linenomath*}
where the functions $k^\pm = k^\pm (\vect{r})$, $\mu^\pm = \mu^\pm(\vect{r}) \ge \epsilon > 0$, $f^\pm = f^\pm (\vect{r})$, $\vect{r} \in \Omega^\pm$, \mylinelabel{rev2:5}\reviewerTwo{and $\alpha = \alpha(\vect{r})$, $\beta = \beta(\vect{r})$, $\vect{r} \in \Gamma$,} are given. We denote by $[q]$ the jump of a scalar quantity $q$ across $\Gamma$, i.e. $[q] = q^+ - q^-$. For simplicity, we impose Dirichlet boundary conditions on the boundary of the computation domain, i.e. $u^+ = g$ on $\partial\Omega$, where $g = g(\vect{r})$ is given.

We discretize the domain $\Omega$ into a uniform rectangular grid of $N_x \times N_y$ points with spatial steps 
\begin{linenomath*}
$$\Delta x = \frac{x_{\max}-x_{\min}}{n_x-1}, \quad \Delta y = \frac{y_{\max}-y_{\min}}{n_y-1}$$
\end{linenomath*}
and associate with each point $\vect{r}_\tm{i,j} = \left( x_i,\, y_j \right) = \left( x_{\min} + (i-1)\Delta x,\, y_{\min} + (j-1) \Delta y \right)$ a finite volume $\cell{i,j} = \left[ x_\tm{i} - \hf \Delta x; x_\tm{i} + \hf \Delta x \right] \times \left[ y_\tm{j} - \hf \Delta y; y_\tm{j} + \hf \Delta y \right]$, $i \in \left[2; N_x-1 \right]$, $j \in \left[ 2; N_y-1 \right]$ (see Fig. \ref{fig::finite_volume}). The Level-Set Method \cite{Osher;Sethian:88:Fronts-propagating-w} is used to describe the irregular interface $\Gamma$. That is, we use a Lipschitz-continuous function $\phi(\vect{r})$ such that $\Omega^+ = \left\{ \vect{r}: \, \phi(\vect{r}) > 0\right\}$, $\Omega^- = \left\{ \vect{r}: \, \phi(\vect{r}) < 0\right\}$ and $\Gamma = \left\{ \vect{r}: \, \phi(\vect{r}) = 0\right\}$.

At the grid points for which the interface $\Gamma$ does not cross the finite volumes, equation \eqref{eq::poisson} is discretized using the standard five-point stencil. Let us consider a point $\vect{r}_\tm{i,j}$ with its finite volume $\cell{i,j}$ crossed by $\Gamma$. Integrating equations \eqref{eq::poisson} over $\cell{i,j}$ and applying the divergence theorem, one gets the following expression:
\begin{linenomath*}
\begin{align*}
\underbrace{
  \sum_{s = +,-}
  \int_{\Omega^s \cap \cell{i,j}}
  k^s u^s
  \diff{\Omega}
}_\text{Linear term}
-
\underbrace{
  \sum_{s = +,-}
  \int_{\Omega^s \cap \partial\cell{i,j}}
  \mu^s \ddna{s}{u^s}
  \diff{\Gamma}
}_\text{Flux between finite volumes}
=
\underbrace{
  \sum_{s = +,-}
  \int_{\Omega^s \cap \cell{i,j}}
  f^s
  \diff{\Omega}
}_\text{Volumetric generation}
+
\underbrace{
  \vphantom{\sum_{s = +,-}}
  \int_{\Gamma \cap \cell{i,j}}
  \jump{ \mu \ddn{u} }
  \diff{\Gamma},
}_\text{Surface generation}
\end{align*}
\end{linenomath*}
where the superscript $^s$ refers to the sign $^\pm$.

Following \cite{Ng;Min;Gibou:09:An-efficient-fluid--, Papac;Gibou;Ratsch:10:Efficient-symmetric-}, that is, approximating the domain integrals by the integrand value multiplied by the corresponding volumes, and estimating the fluxes between cells using values at nearest-neighbor grid points and central difference formulas, one obtains:
\begin{linenomath*}
\begin{align}
\begin{aligned}
\sum_{s = +,-}
k_\tm{i,j}^s u^s_{i,j}
|\cell{i,j}^s|
-
\sum_{s = +,-}
\Bigg(
\mu^s_\tm{i-\hf,j} A_\tm{i-\hf,j}^s \frac{ u^s_\tm{i-1,j} - u^s_\tm{i,j}}{\Delta x} &+
\mu^s_\tm{i+\hf,j} A_\tm{i+\hf,j}^s \frac{ u^s_\tm{i+1,j} - u^s_\tm{i,j}}{\Delta x} +
\\
\mu^s_\tm{i,j-\hf} A_\tm{i,j-\hf}^s \frac{ u^s_\tm{i,j-1} - u^s_\tm{i,j}}{\Delta y} &+
\mu^s_\tm{i,j+\hf} A_\tm{i,j+\hf}^s \frac{ u^s_\tm{i,j+1} - u^s_\tm{i,j}}{\Delta y}
\Bigg)
\\
=
\sum_{s = +,-}
f_\tm{i,j}^s
|\cell{i,j}^{s}|
&+
\int_{\Gamma \cap \cell{i,j}}
\beta
\diff{\Gamma} + \Oof{h^\mathcal{D}},
\end{aligned}
\label{eq::discretization}
\end{align}
\end{linenomath*}
where $\mathcal{D}$ is the problem dimensionality, $h = \max(\Delta x, \Delta y)$,  $|\cell{i,j}^\pm|$ denotes the volume of $\cell{i,j} \cap \Omega^\pm$, $u_\tm{i,j} = u(\vect{r}_\tm{i,j})$, $A^\pm_\tm{i\pm\hf, j}$ and $A^\pm_\tm{i, j\pm\hf}$ are face areas of $\cell{i,j}^\pm$ in the $x$- and $y$-directions, respectively. To compute the boundary and domain integrals required by the proposed discretization, we use the geometric reconstruction approach from \cite{Min;Gibou:07:Geometric-integratio}. In case when an immersed interface is only piece-wise smooth the method from \cite{Bochkov;Gibou:19:Solving-the-Poisson-} can be used.

The discretization given by equation \eqref{eq::discretization} requires that both values of $u^-$ and $u^+$ be available at grid points with a control volume crossed by $\Gamma$, \mylinelabel{rev2:unknowns}\reviewerTwo{thus, one more equation is required at such grid points for the system of equations to be uniquely invertible. We derive the additional equation based on the jump conditions \eqref{eq::jump_value}-\eqref{eq::jump_flux} and Taylor expansions of $u^\pm$ in the normal to the interface direction. Moreover, equations derived in this way can be used to express $u^\pm_\tm{i,j}$ near the interface as a function of $u^\mp_\tm{i,j}$ and the jump conditions \eqref{eq::jump_value}-\eqref{eq::jump_flux}. As a result, it enables us to eliminate the additional degrees of freedom, i.e., reduce the system's size back to $N_x \times N_y$, and make its structure more homogeneous (all of the equations in the linear system are of the same type). This is expected to make the system of equations less difficult to invert by black-box linear solvers.}
We select the $N_x \times N_y$ unknowns to solve for as:
\begin{linenomath*}
\begin{align}
  \label{eq::unknowns}
  u_\tm{i,j} = 
  \begin{cases}
    u^+_\tm{i,j}, \quad \vect{r}_\tm{i,j} \in \Omega^+,\\
    u^-_\tm{i,j}, \quad \vect{r}_\tm{i,j} \in \Omega^-.
  \end{cases}
\end{align} 
\end{linenomath*}
We then develop formulas to express $u^+_\tm{i,j}$ for $\vect{r}_\tm{i,j} \in \Omega^-$ and for $u^-_\tm{i,j}$ for $\vect{r}_\tm{i,j} \in \Omega^+$ as a function of the unknowns $u_\tm{i,j}$. This is described next.

Consider a grid point $\vect{r}_\tm{i,j}$ near the interface $\Gamma$ and its projection, $\vect{r}_\tm{i,j}^{\text{pr}}$, onto the interface (see Fig. \ref{fig::projection}). Taylor expansion relates the values of $u^\pm$ at $\vect{r}_\tm{i,j}$ and $\vect{r}_\tm{i,j}^{\text{pr}}$ as:
\begin{linenomath*}
\begin{align}
  \label{eq::taylor}
  u^\pm_\tm{i,j} = u^\pm (\vect{r}^{\text{pr}}_\tm{i,j}) + \delta_\tm{i,j} \ddn{u^\pm} (\vect{r}^{\text{pr}}_\tm{i,j}) + \Oof{h^2},
\end{align}
\end{linenomath*}
where $\delta_\tm{i,j}$ is the signed distance from $\vect{r}_\tm{i,j}$ to $\vect{r}_\tm{i,j}^{\text{pr}}$ ($\pm \delta_\tm{i,j} > 0$ if $\vect{r}_\tm{i,j} \in \Omega^\pm$). The geometrical quantities $\vect{n} (\vect{r}_\tm{i,j}^{\text{pr}})$, $\vect{r}_\tm{i,j}^{\text{pr}}$ and $\delta_\tm{i,j}$ are estimated from the level-set function as:
\begin{linenomath*}
\begin{eqnarray*}
  \vect{n} (\vect{r}_\tm{i,j}^{\text{pr}}) &=& \vect{n}_\tm{i,j} + \Oof{h} \quad \textrm{ where } \quad
  \vect{n}_\tm{i,j} = \dfrac{\nabla \phi(\vect{r}_\tm{i,j})}{\abs{ \nabla \phi(\vect{r}_\tm{i,j}) }}, \\
  \vect{r}_\tm{i,j}^{\text{pr}} &=& \vect{r}_\tm{i,j} - \delta_\tm{i,j} \vect{n}_\tm{i,j} + \Oof{h^2}, \\
  \delta_\tm{i,j} &=& \frac{\phi(\vect{r}_\tm{i,j})}{\abs{ \nabla \phi(\vect{r}_\tm{i,j}) }} + \Oof{h^2}. \\ 
\end{eqnarray*}
\end{linenomath*}
Subtracting $u^-_\tm{i,j}$ from $u^+_\tm{i,j}$ given in \eqref{eq::taylor} and taking into account the jump condition \eqref{eq::jump_value} one obtains:
\begin{linenomath*}
\begin{align*}
  u_\tm{i,j}^+ - u_\tm{i,j}^- = \alpha (\vect{r}^{\text{pr}}_\tm{i,j}) + \delta_\tm{i,j} 
  \left( \ddn{u^+}(\vect{r}_\tm{i,j}^{\text{pr}}) - \ddn{u^-}(\vect{r}_\tm{i,j}^{\text{pr}}) \right) + \Oof{h^2}.
\end{align*}
\end{linenomath*}
Furthermore, eliminating either $\ddn{u^+}(\vect{r}_\tm{i,j}^{\text{pr}})$ or $\ddn{u^-}(\vect{r}_\tm{i,j}^{\text{pr}})$ in the above expression using the jump condition \eqref{eq::jump_flux} results in the following two equations:
\begin{linenomath*}
\begin{align}
  \label{eq::idea}
  u_\tm{i,j}^+ - u_\tm{i,j}^- = 
  \begin{cases}
    \displaystyle
    \alpha (\vect{r}^{\text{pr}}_\tm{i,j}) + \delta_\tm{i,j} \frac{\beta (\vect{r}^{\text{pr}}_\tm{i,j})}{\mu^+(\vect{r}^{\text{pr}}_\tm{i,j})} - \delta_\tm{i,j} \frac{\mu^+(\vect{r}^{\text{pr}}_\tm{i,j}) - \mu^-(\vect{r}^{\text{pr}}_\tm{i,j})}{\mu^+(\vect{r}^{\text{pr}}_\tm{i,j})} \ddn{u^-}(\vect{r}_\tm{i,j}^{\text{pr}}) \\
    \displaystyle
    \alpha (\vect{r}^{\text{pr}}_\tm{i,j}) + \delta_\tm{i,j} \frac{\beta(\vect{r}^{\text{pr}}_\tm{i,j})}{\mu^-(\vect{r}^{\text{pr}}_\tm{i,j})} - \delta_\tm{i,j} \frac{\mu^+(\vect{r}^{\text{pr}}_\tm{i,j}) - \mu^-(\vect{r}^{\text{pr}}_\tm{i,j})}{\mu^-(\vect{r}^{\text{pr}}_\tm{i,j})} \ddn{u^+}(\vect{r}_\tm{i,j}^{\text{pr}})
  \end{cases}
  + \Oof{h^2}.
\end{align}
\end{linenomath*}
If one approximates either $\ddn{u^+}(\vect{r}_\tm{i,j}^{\text{pr}})$ or $\ddn{u^-}(\vect{r}_\tm{i,j}^{\text{pr}})$ using $u_\tm{i,j}^\pm$ and $\set{u_\tm{p,q}, p\in[1,N_x], q\in[1,N_y]}{}{}$, then these formulas can be used to eliminate additional degrees of freedom. \mylinelabel{rev2:deriv1}\reviewerTwo{A straightforward way to do that is to use a suitable local interpolants $u^\pm_I=u^\pm_I(\vect{r})$ of unknown functions $u^\pm$ as: 
\begin{linenomath*}
\begin{align}\label{eq::norm_deriv}
  \ddn{u^\pm}(\vect{r}_\tm{i,j}^{\text{pr}}) 
  = \vect{n}(\vect{r}_\tm{i,j}^{\text{pr}})  \nabla u^\pm_I (\vect{r}_\tm{i,j}^{\text{pr}}).
\end{align}
\end{linenomath*} 
Specifically, in this work we use the linear interpolation:}
\begin{linenomath*}\mylinelabel{rev2:interpolant}
\begin{align}\label{eq::interpolant}
  u^\pm_I(\vect{r}) = u^\pm_\tm{i,j} + \left(\vect{r} - \vect{r}_\tm{i,j}\right)^T \left( \nabla u^\pm \right)_\tm{i,j}  + \Oof{h^2},
\end{align}
\end{linenomath*}
where the gradient $\left( \nabla u^\pm \right)_\tm{i,j}$ is found as the least-square solution satisfying the constraints:
\begin{linenomath*}
\begin{align*}
  u_\tm{i+p,j+q} = u^\pm_{i,j} + \left(\vect{r}_\tm{i+p,j+q} - \vect{r}_\tm{i,j}\right)^T \left( \nabla u^\pm \right)_\tm{i,j} , \quad (p,q) \in N^\pm_\tm{i,j}.
\end{align*}
\end{linenomath*}
$N^\pm_\tm{i,j}$ denotes the set of neighboring grid points of $\vect{r}_\tm{i,j}$, lying in the region $\Omega^\pm$, that is:
\begin{linenomath*}
\begin{align*}
N^\pm_\tm{i,j} &= \left\{ (p,q):\quad  p = -1,0,1,\quad q = -1, 0, 1, \quad (p,q) \neq (0, 0), \quad  \vect{r}_\tm{i+p, j+q} \in \Omega^\pm \right\}.
\end{align*}
\end{linenomath*}
\mylinelabel{rev2:deriv2}\reviewerTwo{In other words, the local linear interpolants are constructed using available values at the nearest-neighbor  grid points of $\vect{r}_\tm{i,j}$ (in Cartesian and diagonal directions).} Note also that $u_\tm{i+p,j+q} = u^\pm_\tm{i+p,j+q}$ if $(p,q) \in N^\pm_\tm{i,j}$.

Thus, the gradient $\left( \nabla u^\pm \right)_\tm{i,j}$ is the least-squares solution of the following linear system:
\begin{linenomath*}
\begin{align*}
  \mat{X}_\tm{i,j} \mat{W}_\tm{i,j}^\pm \left(\nabla u^\pm \right)_\tm{i,j} = \mat{W}_\tm{i,j}^\pm
  \begin{pmatrix}
    u_{i-1,j-1} - u^\pm_{i,j} \\
    u_{i, j-1} - u^\pm_{i,j} \\
    \ldots \\
    u_{i+1, j+1} - u^\pm_{i,j}
  \end{pmatrix},
\end{align*}
\end{linenomath*}
that is:
\begin{linenomath*}
\begin{align*}
  \left(\nabla u^\pm \right)_\tm{i,j} = 
  \mat{D}_\tm{i,j}^\pm
    \begin{pmatrix}
      u_{i-1,j-1} - u^\pm_{i,j} \\
      u_{i, j-1} - u^\pm_{i,j} \\
      \ldots \\
      u_{i+1, j+1} - u^\pm_{i,j}
    \end{pmatrix},
  \quad
  \mat{D}_\tm{i,j}^\pm
  =\left( \mat{X}_\tm{i,j}^T \mat{W}^\pm_\tm{i,j} \mat{X}_\tm{i,j} \right)^{-1} \left( \mat{W}^\pm_\tm{i,j} \mat{X}_\tm{i,j} \right)^T,
\end{align*}
\end{linenomath*}
where the $3^\mathcal{D} \times \mathcal{D}$ and $3^\mathcal{D} \times 3^\mathcal{D}$ matrices $\mat{X}_\tm{i,j}$ and $\mat{W}_\tm{i,j}$ are given by:
\begin{linenomath*}
\begin{align*}
  \mat{X}_\tm{i,j} =
  \begin{pmatrix}
    \left( \vect{r}_\tm{i-1, j-1} - \vect{r}_\tm{i,j} \right)^T \\
    \left( \vect{r}_\tm{i, j-1} - \vect{r}_\tm{i,j} \right)^T \\
    \ldots \\
    \left( \vect{r}_\tm{i+1, j+1} - \vect{r}_\tm{i,j} \right)^T    
  \end{pmatrix}
\quad \textrm{ and } \quad
  \mat{W}_\tm{i,j}^\pm = 
  \begin{pmatrix}
    \omega^\pm_\tm{i,j}(-1,-1) & \, & \, & \, \\
    \, & \omega^\pm_\tm{i,j}(0,-1) & \, & \, \\
        \, & \, & \ddots & \, \\
    \, & \, & \, & \omega^\pm_\tm{i,j}(1,1)
  \end{pmatrix},
\end{align*}
\end{linenomath*} with
\begin{linenomath*}
\begin{align*}
  \omega_\tm{i,j}^\pm(p,q) = 
  \begin{cases}
    1,\, (p,q) \in N^\pm_\tm{i,j} \\
    0,\, (p,q) \in N^\mp_\tm{i,j}
  \end{cases}.
\end{align*}
\end{linenomath*}

\mylinelabel{rev2:deriv3}\reviewerTwo{Substitution of $u^\pm_I(\vect{r})$ into $\eqref{eq::norm_deriv}$
yields approximations of $\ddn{u^\pm}(\vect{r}_\tm{i,j}^{\text{pr}})$ as linear combinations of $\set{u_\tm{i+p,j+q}, p=-1,0,1, q=-1,0,1}{}{}$}. 
Specifically, let us write the $\mathcal{D} \times 3^{\mathcal{D}}$ matrix $\mat{D}^\pm_\tm{i,j}$ as:
\begin{linenomath*}
\begin{align*}
  \mat{D}^\pm_\tm{i,j} =
  \begin{pmatrix}
    \vect{d}_{i,j,-1,-1}^\pm &
    \vect{d}_{i,j, 0,-1}^\pm &
    \cdots &
    \vect{d}_{i,j, 1, 1}^\pm
  \end{pmatrix},
\end{align*}
\end{linenomath*}
\mylinelabel{rev2:ds}\reviewerTwo{where vectors $\vect{d}_{i,j,-1,-1}^\pm$, $\ldots$, $\vect{d}_{i,j, 1, 1}^\pm$ represent the columns of the matrix $\mat{D}^\pm_\tm{i,j}$.}
Then the normal derivative can be expressed as:
\begin{linenomath*}
\begin{align}
  \ddn{u^\pm}(\vect{r}_\tm{i,j}^{\text{pr}}) &= c^\pm_{i,j} u^\pm_\tm{i,j} + \sum_{(p,q)\in N^\pm_\tm{i,j}} c^\pm_\tm{i,j,p,q} u_\tm{i+p,j+q} + \Oof{h}, \label{eq::normalDerivative}
\end{align}
\end{linenomath*}
where the coefficients are given by:
\begin{linenomath*}
\begin{align*}
  c^\pm_\tm{i,j,p,q} = \vect{n}_\tm{i,j}^T \vect{d}_{i,j,p,q}^\pm, 
  \quad (p,q) \in N^\pm_\tm{i,j},
  \quad \textrm{and} \quad
  c^\pm_\tm{i,j} = - \sum_{(p,q)\in N^\pm_\tm{i,j}} c^\pm_\tm{i,j,p,q}.
\end{align*}
\end{linenomath*}
Substitution of \eqref{eq::normalDerivative} into \eqref{eq::idea} produces formulas expressing $u^+_\tm{i,j}$ and $u^-_\tm{i,j}$ in terms of the selected $N_x \times N_y$ unknowns $\set{u_\tm{p,q}, p\in[1,N_x], q\in[1,N_y]}{}{}$. Combining them with the definition \eqref{eq::unknowns}, we get the following two sets of rules (which are $\Oof{h^2}$ \mylinelabel{rev2:o2inu}\reviewerTwo{accurate in the value of $u$}): \mylinelabel{rev2:rules}\reviewerTwo{one is based on approximating $\ddn{u^+}(\vect{r}_\tm{i,j}^{\text{pr}})$:
\begin{linenomath*}
\begin{align*}
  u^-_\tm{i,j} &= 
    \begin{cases}
      u_\tm{i,j},\, &\vect{r}_{i,j} \in \Omega^-,
      \\
        u_\tm{i,j}
        - \alpha 
        - \delta_\tm{i,j} 
          \frac{\beta }{\mu^+  }
          -
          \delta_\tm{i,j}
          \frac{\jump{\mu}}{\mu^+ }
          \frac{
          \left(
          c_\tm{i,j}^- 
          \left( u_\tm{i,j} - \alpha - \delta_\tm{i,j} \frac{\beta}{\mu^+ } \right)
          + \sum_{(p,q)\in N^-_\tm{i,j}} c^-_\tm{i,j,p,q} u_\tm{i+p,j+q}
          \right)
          }
          {\left(1 - \delta_\tm{i,j} \frac{\jump{\mu}}{\mu^+} c_\tm{i,j}^-\right)}
          ,\, &\vect{r}_{i,j} \in \Omega^+,
    \end{cases}
  \\
  u^+_\tm{i,j} &=
    \begin{cases}
       u_\tm{i,j}
  + \alpha
  + \delta_\tm{i,j} 
    \frac{\beta}{\mu^+}
    -
    \delta_\tm{i,j}
    \frac{\jump{\mu}}{\mu^+} 
    \left( c_\tm{i,j}^- u_\tm{i,j}
    +
    \sum_{(p,q)\in N^-_\tm{i,j}} c^-_\tm{i,j,p,q} u_\tm{i+p,j+q} 
    \right),\, &\vect{r}_{i,j} \in \Omega^-,
    \\
    u_\tm{i,j},\, &\vect{r}_{i,j} \in \Omega^+, 
  \end{cases}
\end{align*}
\end{linenomath*}
while the other one is based on approximating $\ddn{u^-}(\vect{r}_\tm{i,j}^{\text{pr}})$:
\begin{linenomath*}
\begin{align*}
  u^-_\tm{i,j} &= 
    \begin{cases}
      u_\tm{i,j},\, &\vect{r}_{i,j} \in \Omega^-,
      \\
      u_\tm{i,j}
      - \alpha 
      - \delta_\tm{i,j} 
      \frac{\beta }{\mu^-  }
      -
      \delta_\tm{i,j}
      \frac{\jump{\mu} }{\mu^-  }
      \left( c_\tm{i,j}^+ u_\tm{i,j} 
      + \sum_{(p,q)\in N^+_\tm{i,j}} c^+_\tm{i,j,p,q} u_\tm{i+p,j+q} \right),\, &\vect{r}_{i,j} \in \Omega^+,
    \end{cases}
  \\
  u^+_\tm{i,j} &=
    \begin{cases}
       u_\tm{i,j} 
  + \alpha
  + \delta_\tm{i,j} 
  \frac{\beta}{\mu^-}
  -
  \delta_\tm{i,j} 
  \frac{\jump{\mu}}{\mu^- }
  \frac{
  \left(c_\tm{i,j}^+ 
  \left( u_\tm{i,j} + \alpha + \delta_\tm{i,j} \frac{\beta}{\mu^- } \right)
  + \sum_{(p,q)\in N^+_\tm{i,j}} c^+_\tm{i,j,p,q} u_\tm{i+p,j+q}\right)
  }
  {\left(1 + \delta_\tm{i,j} \frac{\jump{\mu}}{\mu^-} c_\tm{i,j}^+\right)}
  ,\, &\vect{r}_{i,j} \in \Omega^-,
    \\
    u_\tm{i,j},\, &\vect{r}_{i,j} \in \Omega^+.
  \end{cases}
\end{align*}
\end{linenomath*}
}

Thus, one has a certain flexibility in constructing the final discretization. For example, one could choose, for each $\vect{r}_\tm{i,j}$, the formula based on approximating $\ddn{u^-}(\vect{r}_\tm{i,j}^{\text{pr}})$ or $\ddn{u^+}(\vect{r}_\tm{i,j}^{\text{pr}})$ depending on the largest number of neighboring points of $\vect{r}_\tm{i,j}$ that are in $\Omega^-$ or in $\Omega^+$ (let us denote this scheme as  \texttt{Random}). However, this choice would ignore the magnitude of the diffusion constants $\mu^-$ and $\mu^+$ and their influence on the condition number of the linear system. To investigate this issue, we consider two additional schemes: the first one (referred to as \texttt{Bias Fast}) uses interpolation in the fast-diffusion region (i.e., if $\mu^- > \mu^+$ then the formula based on $\ddn{u^-}(\vect{r}_\tm{i,j}^{\text{pr}})$ is used); the second scheme (referred to as \texttt{Bias Slow}) uses interpolation in the slow-diffusion region (i.e., if $\mu^- > \mu^+$ then we use the formula based on $\ddn{u^+}(\vect{r}_\tm{i,j}^{\text{pr}})$). 

\vspace{.25cm}
\noindent\textbf{Remarks}:
\begin{itemize}
\item In the limiting cases $\frac{\mu^-}{\mu^+} \rightarrow \infty$ or $\frac{\mu^-}{\mu^+} \rightarrow 0$, only the scheme \texttt{Bias Slow} remains well defined, thus, we expect it to perform the best and be well-conditioned for any ratio of diffusion coefficients. We will illustrate in section \ref{sec::NumericalTest} that only the scheme \texttt{Bias Slow} produces a condition number that is bounded. \mylinelabel{ours:compare}\ours{This is consistent with the results reported in \cite{Xu:12:An-iterative-two-flu} describing a variant of the augmented immersed interface method.}

\item In the limiting case $\frac{\mu^-}{\mu^+} \equiv 1$, the three schemes coincide. Moreover, the matrix associated with the resulting linear system is the same as for the case when no interface is present (that is, as for the standard five-point stencil in 2D) and only the right-hand is changed to account for jump conditions.

\item The truncation error is the same for all three schemes. Therefore, we expect them to have similar accuracies. Specifically, the truncation error\footnote{After scaling the resulting discretization by the cell volume $\Oof{h^\mathcal{D}}$ to account for the integration of the PDEs over a finite volume.} is $\Oof{h^2}$ for grid points away from the immersed interface and $\Oof{1}$ for cells crossed by the interface. Following the results of \mylinelabel{rev2:refs}\cite{Johansen;Colella:98:A-Cartesian-Grid-Emb,Gibou;Fedkiw;Cheng;etal:02:A-second-order-accur,Schwartz;Barad;Colella;etal:06:A-Cartesian-grid-emb,Chen;Min;Gibou:07:A-supra-convergent-f,Ng;Chen;Min;etal:09:Guidelines-for-Poiss,Papac;Gibou;Ratsch:10:Efficient-symmetric-,Gallinato;Poignard:15:Superconvergent-Cart, Gallinato;Poignard:17:Superconvergent-seco,Bochkov;Gibou:19:Solving-the-Poisson-}, we expect the schemes to produce second-order accurate numerical solutions with first-order accurate gradients.

\item \mylinelabel{rev2:accuracy}\reviewerTwo{The truncation error can be improved to be $\Oof{h}$ for cells crossed by the interface by making the following changes in the discretization scheme: 1) Estimate fluxes between cells at the centroids of cell faces using linear interpolation as done, for example, in \cite{Johansen;Colella:98:A-Cartesian-Grid-Emb, Bochkov;Gibou:19:Solving-the-Poisson-}; 2) Retain the quadratic term in Taylor expansion \eqref{eq::taylor}; 3) Use a quadratic interpolant (instead of linear \eqref{eq::interpolant}) to approximate $\ddn{u}^\pm$ at projection points. This is expected to increase the accuracy of solution gradients to second order. We leave this investigation to future work.
}

\item In general the case where \mylinelabel{rev2:4}\reviewerTwo{$\mu^+ \neq \mu^-$}, the resulting linear system is nonsymmetric. In the worst case scenario the computational stencil involves nearest neighbors (both in Cartesian and diagonal directions) of the standard five-point stencil as illustrated in Fig. \ref{fig::stencil}. An example of the matrix associated with the resulting linear system structure is shown in Fig. \ref{fig::matrix}.

\begin{figure}[h]
  \centering
  \begin{subfigure}[b]{0.3\textwidth}
    \centering
    \includegraphics[width=0.9\textwidth]{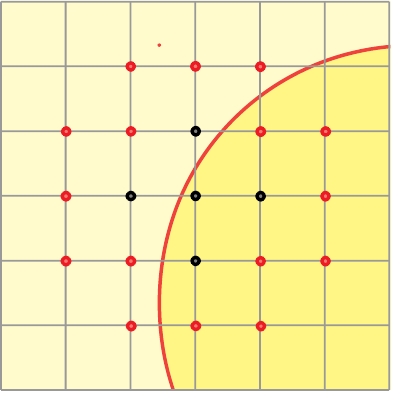}
    \caption{}
    \label{fig::stencil}
  \end{subfigure}
  \begin{subfigure}[b]{0.3\textwidth}
    \centering
    \includegraphics[width=0.9\textwidth]{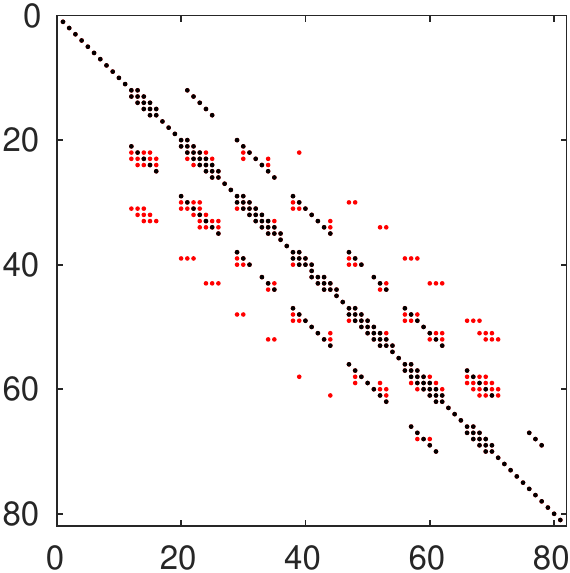}
    \caption{}
    \label{fig::matrix}
  \end{subfigure}
  \caption{(a) Computational stencil (the red color indicates additional grid points used in case $\mu^+ \neq \mu^-$). (b) Matrix structure of the resulting linear system in case of two-dimensional example from Sec. \ref{sec::2dTests} on a $8^2$ grid (the red color indicates additional elements in case $\mu^+ \neq \mu^-$). }
\end{figure}

\end{itemize}

\section{Numerical tests} \label{sec::NumericalTest}
To numerically illustrate the properties of the proposed schemes, we study three characteristics: the order of accuracy of the numerical solution in the $L^\infty$-norm, the order of accuracy of the numerical gradients in the $L^\infty$-norm, and the condition number of the linear system, estimated by the MATLAB \texttt{condest} function. We consider two tests: the first one, the \texttt{convergence test}, studies the dependence of those three characteristics on the grid resolution. The second one, the \texttt{conditioning test}, focuses on the dependence of the three characteristics on the  ratio, $\frac{\mu^-}{\mu^+}$, of the diffusion coefficients. We perform both tests in two and three spatial dimensions. In all the examples, we use the implementation of the BiCGStab algorithm provided by PETSc \cite{Balay;Brown;Buschelman;etal:12:PETSc-Web-page} with the Hypre preconditioner \cite{Falgout;Yang:02:Hypre:-A-library-of-}.

\subsection{Two-dimensional case}\label{sec::2dTests}
\begin{figure}[!h]
  \centering 
  \begin{subfigure}[b]{0.3\textwidth}
    \centering 
    \includegraphics[width=0.9\textwidth]{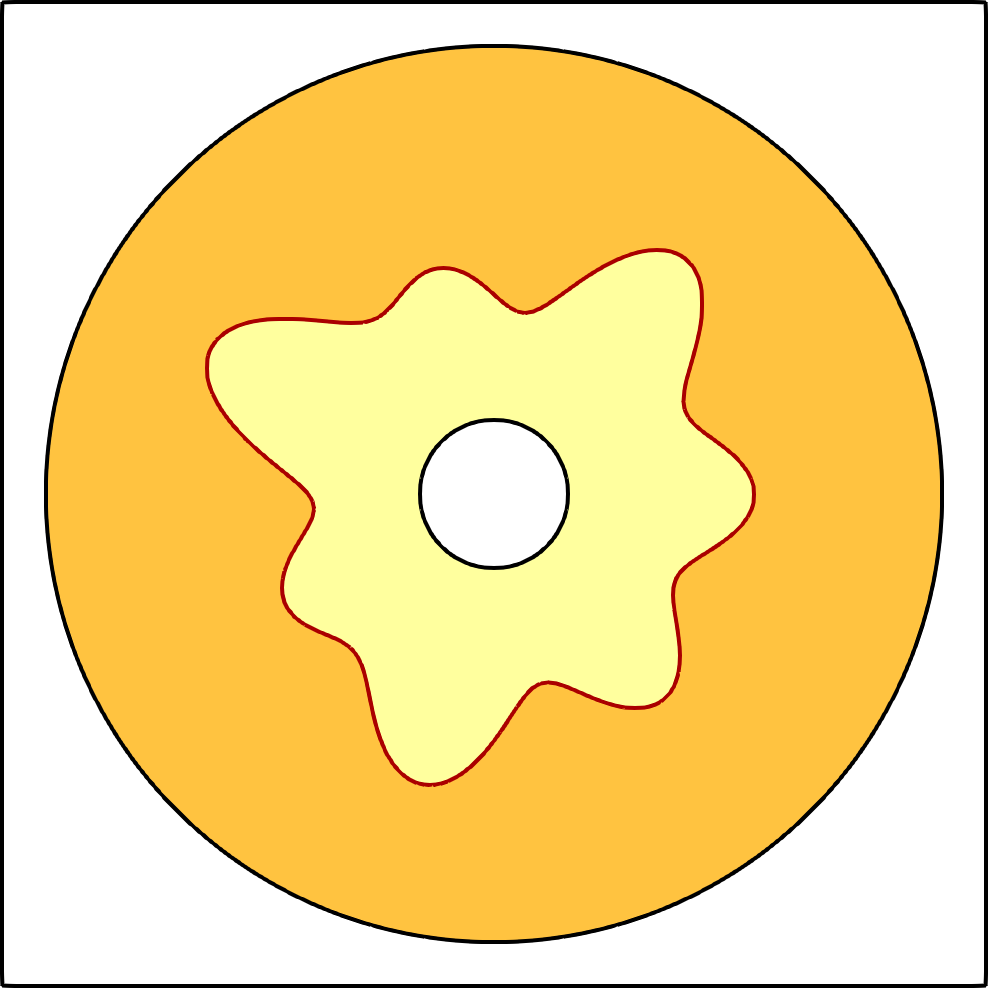}
      \caption{}
      \label{fig::2d::geometry::domain}
  \end{subfigure}
  \begin{subfigure}[b]{0.3\textwidth}
    \centering 
    \includegraphics[width=0.95\textwidth]{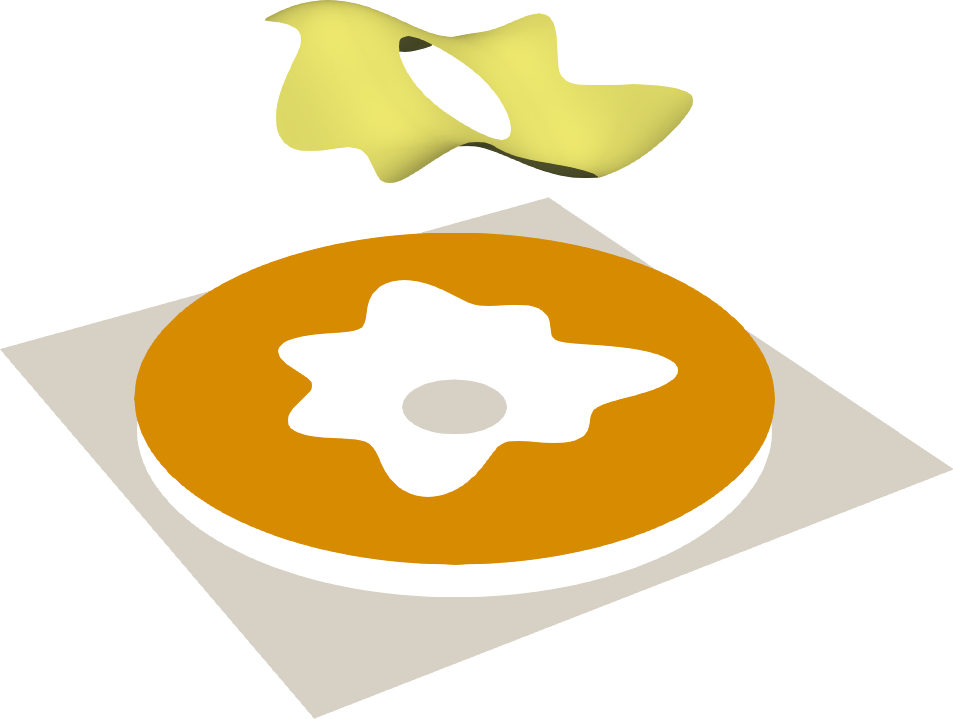}
      \caption{}
      \label{fig::2d::geometry::diffusivity}
  \end{subfigure}
  \begin{subfigure}[b]{0.3\textwidth}
    \centering 
    \includegraphics[width=0.95\textwidth]{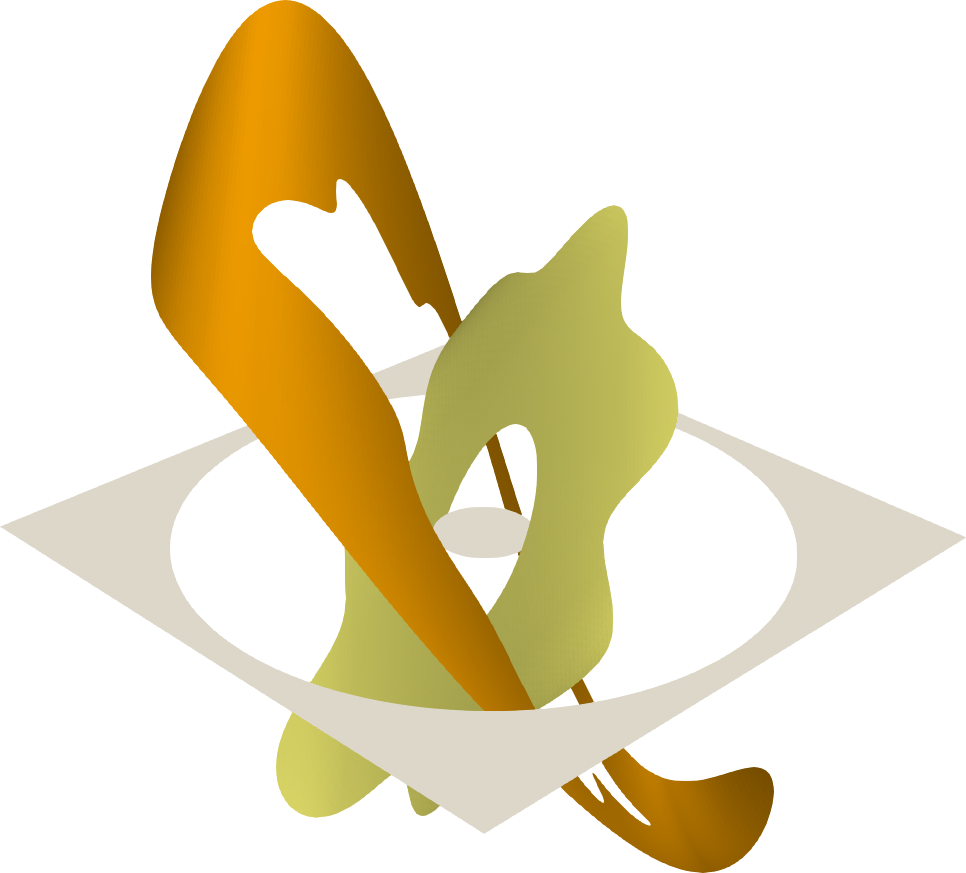}
      \caption{}
      \label{fig::2d::geometry::solution}
  \end{subfigure}
  \caption{\reviewerTwo{(a) Problem geometry. (b) Diffusion coefficients (scaled by $0.1$ for visualization). (c) Numerical solution on a $256^2$ grid.}}
  \label{fig::2d::geometry}
\end{figure}

\begin{figure}[!h]
  \centering 
  \begin{subfigure}[b]{0.32\textwidth}
    \centering 
    \includegraphics[width=0.95\textwidth]{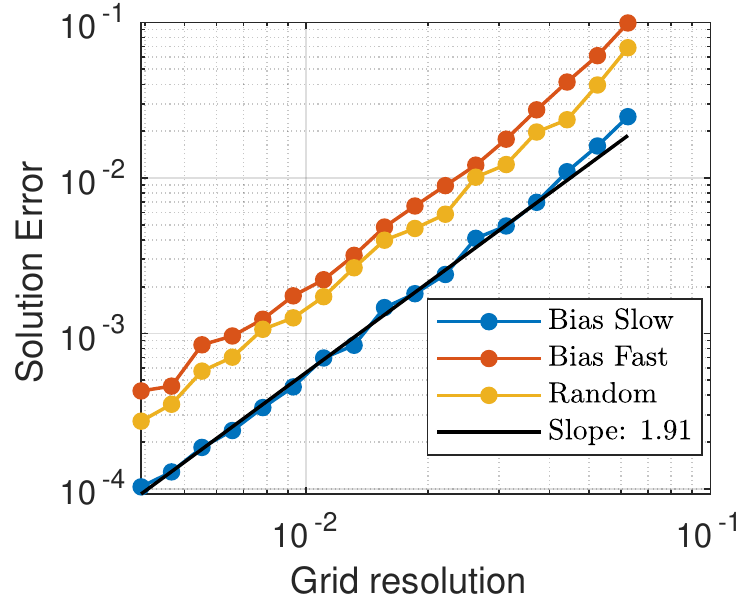}
  \end{subfigure}
  \begin{subfigure}[b]{0.32\textwidth}
    \centering 
    \includegraphics[width=0.95\textwidth]{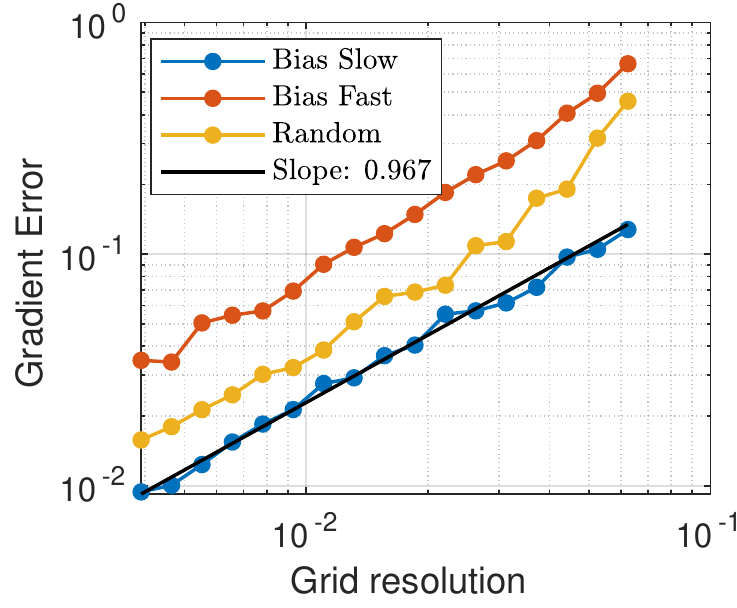}
  \end{subfigure}
  \begin{subfigure}[b]{0.32\textwidth}
    \centering 
    \includegraphics[width=0.95\textwidth]{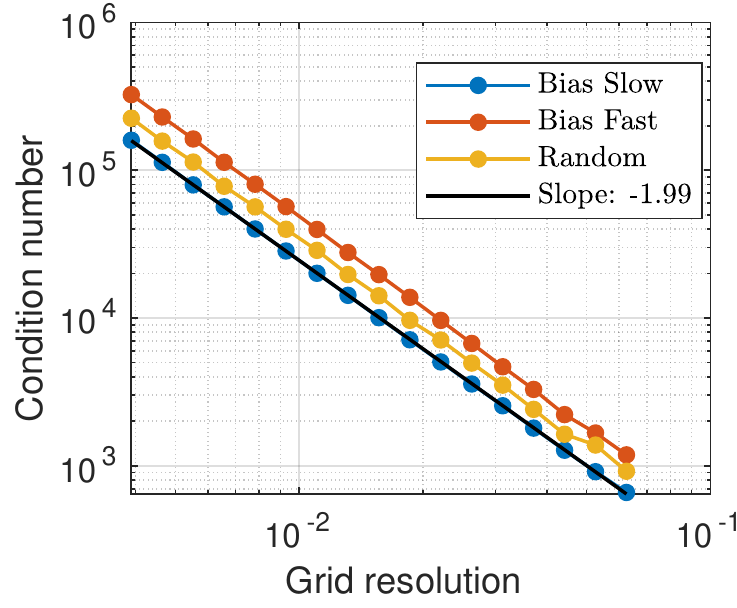}
  \end{subfigure}
  \caption{\texttt{convergence test} in two spatial dimensions (each data point represents the maximum value among $10\times 10 = 100$ different relative placements of an immersed interface on the computational grid).}
  \label{fig::2d::accuracy}
\end{figure}

\begin{figure}[!h]
  \centering 
  \begin{subfigure}[b]{0.32\textwidth}
    \centering 
    \includegraphics[width=0.95\textwidth]{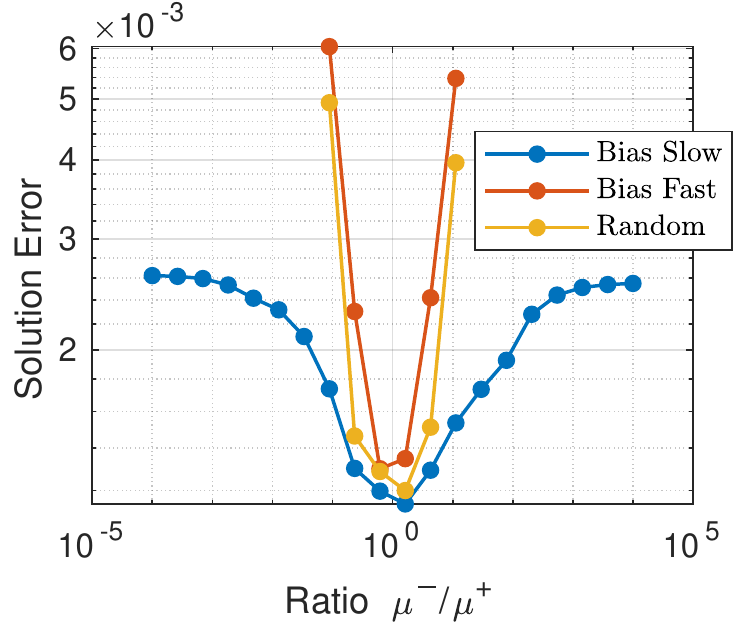}
  \end{subfigure}
  \begin{subfigure}[b]{0.32\textwidth}
    \centering 
    \includegraphics[width=0.95\textwidth]{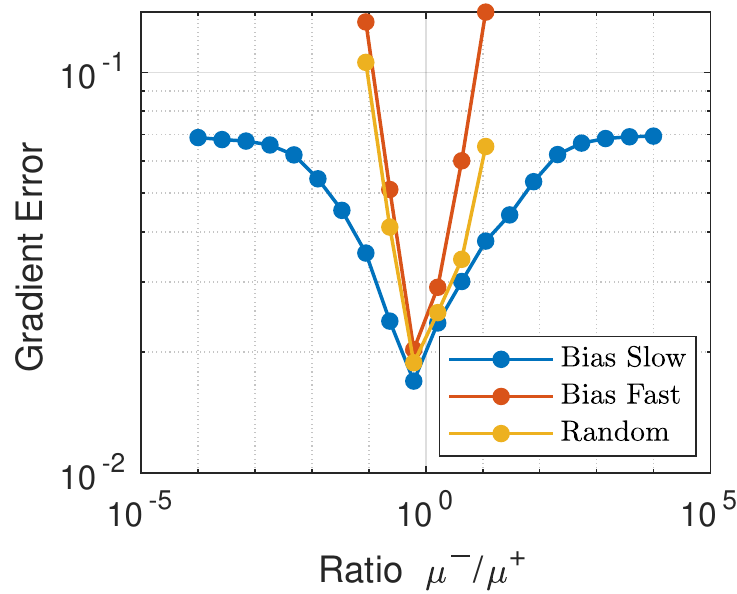}
  \end{subfigure}
  \begin{subfigure}[b]{0.32\textwidth}
    \centering 
    \includegraphics[width=0.95\textwidth]{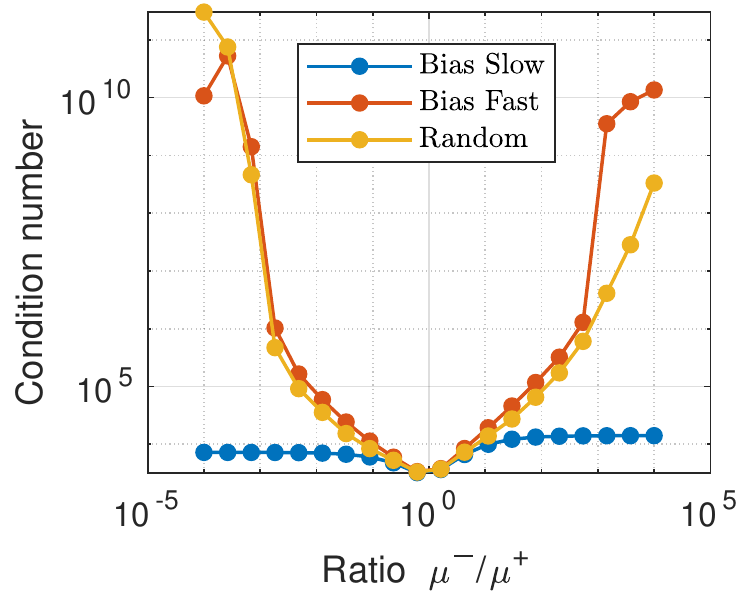}
  \end{subfigure}
  \caption{\texttt{conditioning test} in two spatial dimensions (each data point represents the maximum value among $10\times 10 = 100$ different relative placements of an immersed interface on the computational grid).}
  \label{fig::2d::conditioning}
\end{figure}

Consider an annular region\footnote{We enclose an immersed interface inside another region in order to be able to obtain results for different placements of the immersed interface and the computational grid without changing the problem statement. On the boundaries of the enclosing region, Dirichlet boundary conditions can be imposed with any of the methods \cite{Gibou;Fedkiw;Cheng;etal:02:A-second-order-accur, Gibou;Fedkiw:05:A-fourth-order-accur, Shortley;Weller:38:Numerical-solution-o}} with inner and outer radii $r_i = 0.151$ and $r_e = 0.911$, and an immersed star-shaped interface (see Fig. \ref{fig::2d::geometry::domain}), described by the following level-set function:
\begin{linenomath*}
\begin{align*}
  \phi(x, y) = \sqrt{x^2 + y^2} - r_0 \left(1 + \sum_{k=1}^{3} \beta_k \cos \left(n_k \left(\arctan \left( \frac{y}{x} \right) - \theta_k\right)\right) \right),
\end{align*}
\end{linenomath*} 
with parameters:
\begin{linenomath*}
\begin{align}\label{eq::level-set-params}
  r_0 = 0.483,
  \quad
  \begin{pmatrix}
    n_1 \\ \beta_1 \\ \theta_1
  \end{pmatrix}
  =
  \begin{pmatrix}
    3 \\ 0.1 \\ 0.5
  \end{pmatrix},  
  \quad
  \begin{pmatrix}
    n_2 \\ \beta_2 \\ \theta_2
  \end{pmatrix}
  =
  \begin{pmatrix}
    4 \\ -0.1 \\ 1.8
  \end{pmatrix}  
  \quad
  \text{and}
  \quad
  \begin{pmatrix}
    n_3 \\ \beta_3 \\ \theta_3
  \end{pmatrix}
  =
  \begin{pmatrix}
    7 \\ 0.15 \\ 0
  \end{pmatrix}.
\end{align}
\end{linenomath*}
Using the method of manufactured solutions, we take the exact solution to be $u^- = \sin(2x) \cos(2y)$ and $u^+ = \left( 16 \left( \frac{y-x}{3} \right)^5 - 20 \left( \frac{y-x}{3} \right)^3 + 5 \left( \frac{y-x}{3} \right) \right) \log \left( x+y+3 \right)$ (see Fig. \ref{fig::2d::geometry::solution}). For the \texttt{convergence test}, we set the diffusion coefficients to $\mu^- = 10 \left( 1 + \frac{1}{5} \cos( 2 \pi (x+y) ) \sin( 2 \pi (x-y) ) \right)$ and $\mu^+ = 1$ (see Fig. \ref{fig::2d::geometry::diffusivity}), and we vary the grid resolution from $2^{-4}$ to $2^{-9}$.  For the \texttt{conditioning test}, we fix the grid resolution at $2^{-6}$ and $\mu^+ = 1$ and vary $\mu^-$ from $10^{-4}$ to $10^{4}$. The results are presented in Fig. \ref{fig::2d::accuracy} and \ref{fig::2d::conditioning}, where each data point represents the maximum value among $10 \times 10 = 100$ different relative placements of the immersed interface on the computational grid (as done in \cite{Bochkov;Gibou:19:Solving-the-Poisson-}). The different placements thus account for cases where the interface defines a control volume that is arbitrarily small or large, relative to an elementary grid cell. Section \ref{sec::Analysis} will draw some conclusions from these results.\mylinelabel{rev2:fig3caption}

\subsection{Three-dimensional case}\label{sec::3dTests}
\begin{figure}[!h]
  \centering 
  \begin{subfigure}[b]{0.4\textwidth}
    \centering 
    \includegraphics[width=0.95\textwidth]{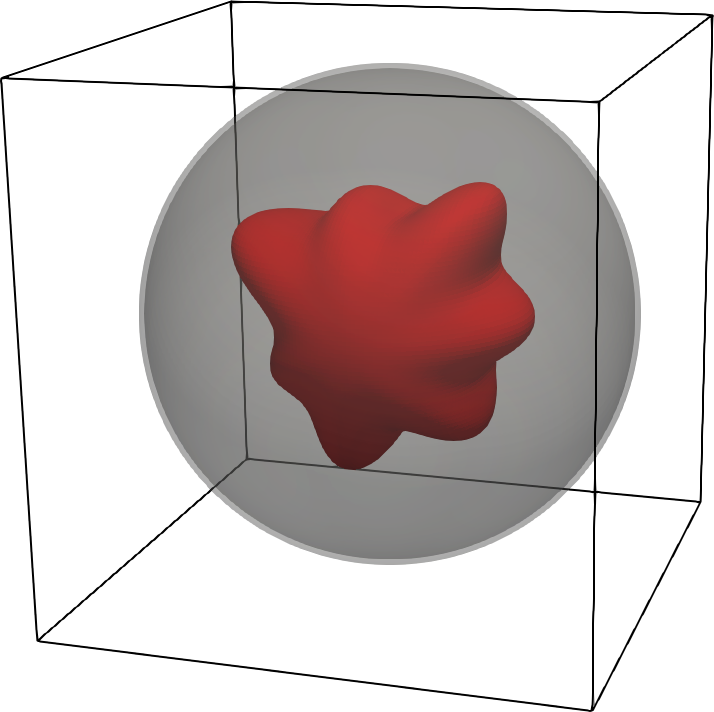}
  \end{subfigure} \hspace{1cm}
  \begin{subfigure}[b]{0.4\textwidth}
    \centering 
    \includegraphics[width=0.95\textwidth]{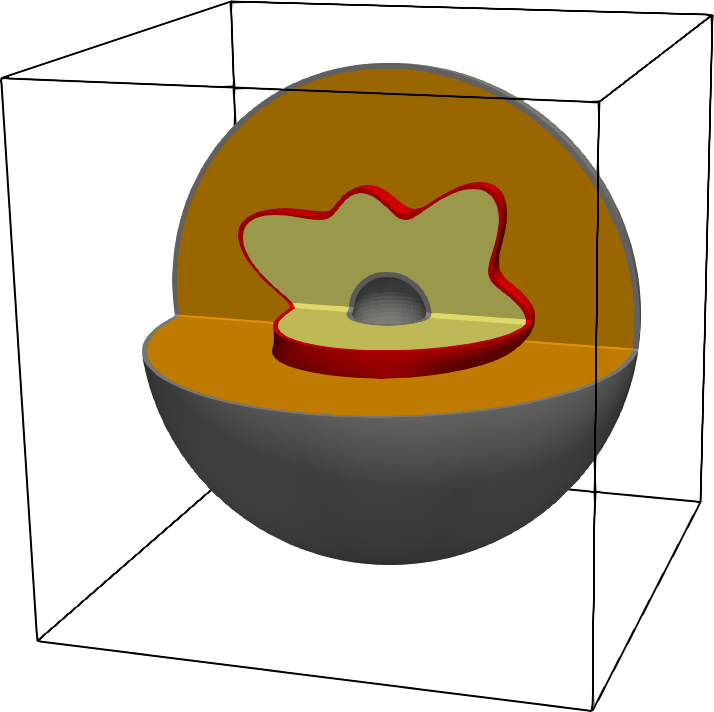}
  \end{subfigure}
  \caption{Illustration of problem geometry in the three-dimensional case. }
  \label{fig::3d::geometry}
\end{figure}

\begin{figure}[!h]
  \centering 
  \begin{subfigure}[b]{0.32\textwidth}
    \centering 
    \includegraphics[width=0.95\textwidth]{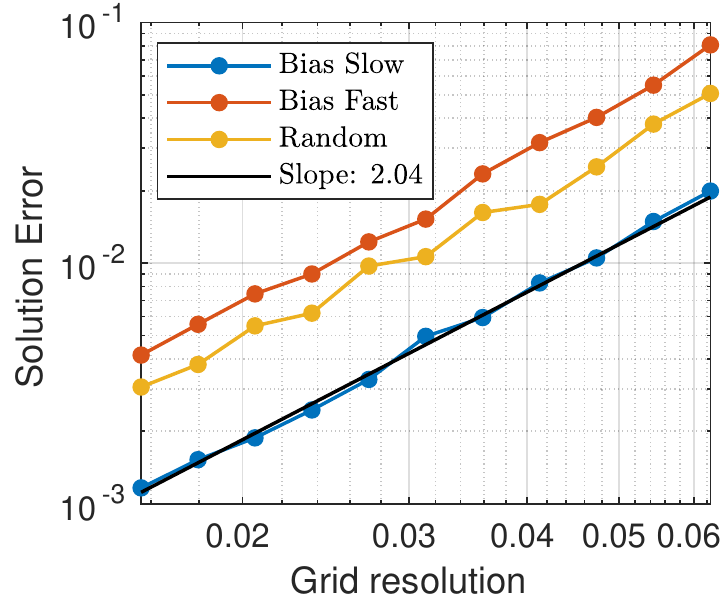}
  \end{subfigure}
  \begin{subfigure}[b]{0.32\textwidth}
    \centering 
    \includegraphics[width=0.95\textwidth]{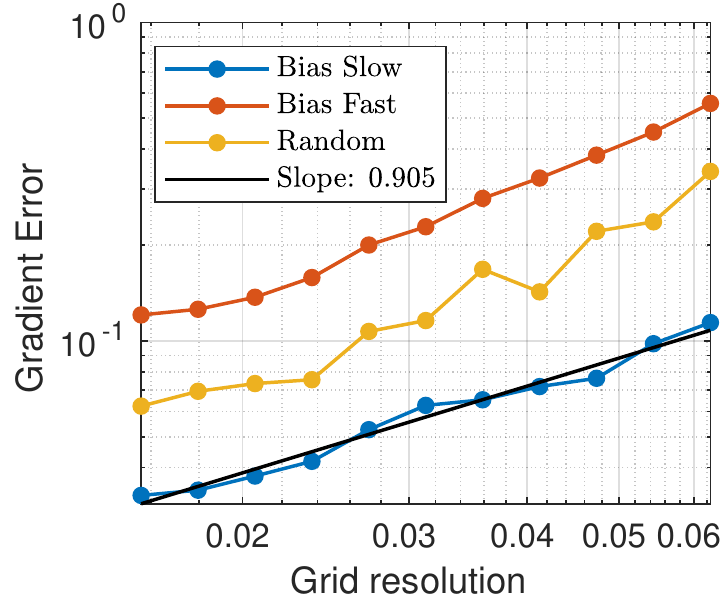}
  \end{subfigure}
  \begin{subfigure}[b]{0.32\textwidth}
    \centering 
    \includegraphics[width=0.95\textwidth]{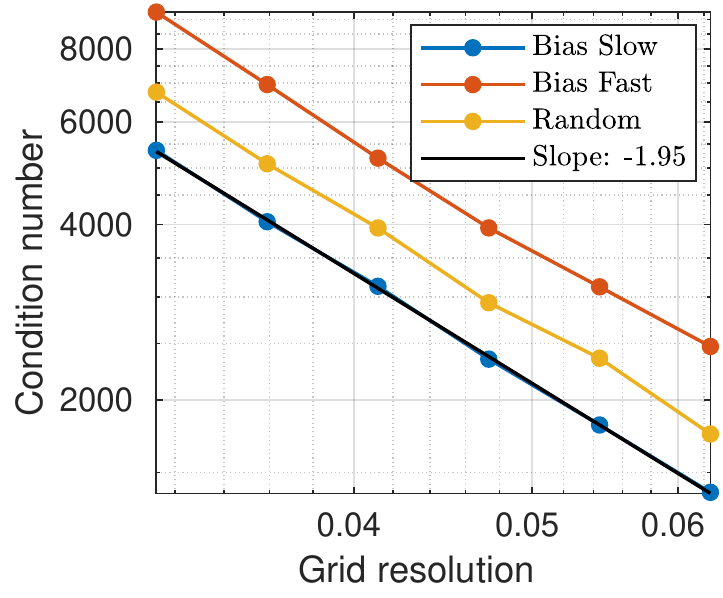}
  \end{subfigure}
  \caption{\texttt{convergence test} in three spatial dimensions (each data point represents maximum value among $5 \times 5 \times 5 = 125$ different relative placements of an immersed interface and the computational grid).}
  \label{fig::3d::accuracy}
\end{figure}

\begin{figure}[!h]
  \centering 
  \begin{subfigure}[b]{0.32\textwidth}
    \centering 
    \includegraphics[width=0.95\textwidth]{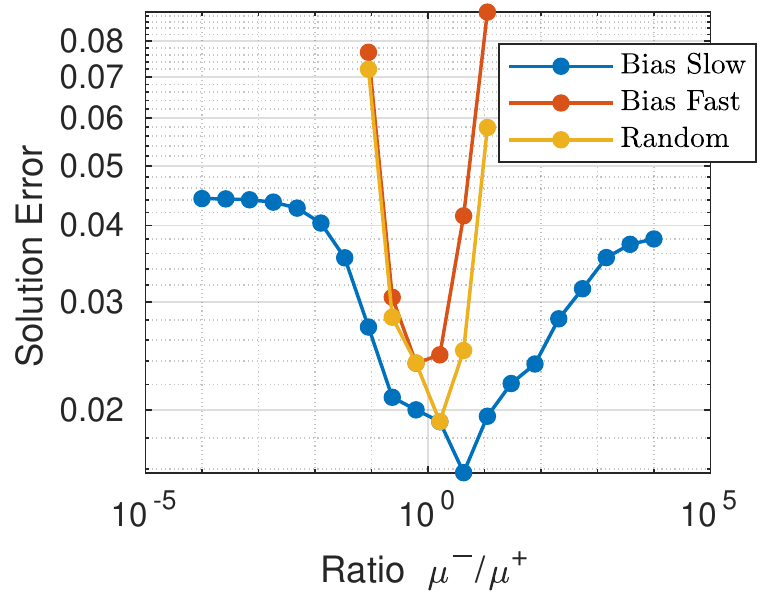}
  \end{subfigure}
  \begin{subfigure}[b]{0.32\textwidth}
    \centering 
    \includegraphics[width=0.95\textwidth]{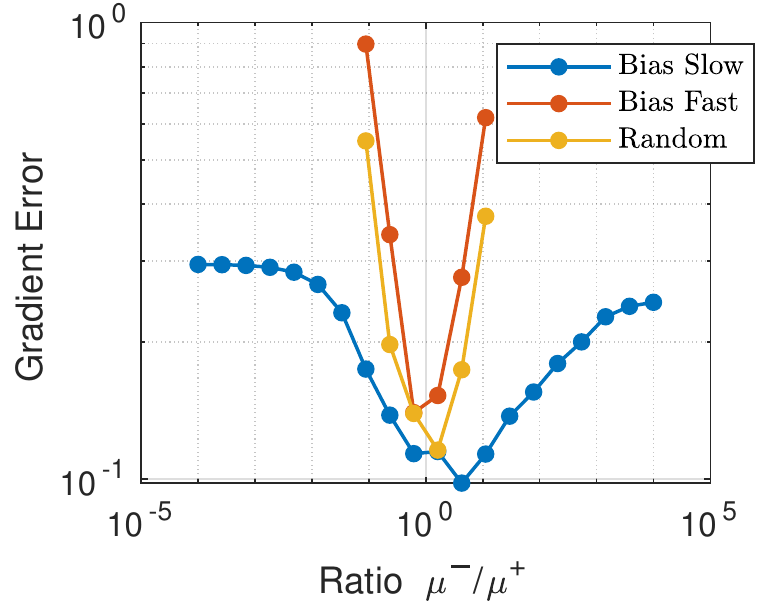}
  \end{subfigure}
  \begin{subfigure}[b]{0.32\textwidth}
    \centering 
    \includegraphics[width=0.95\textwidth]{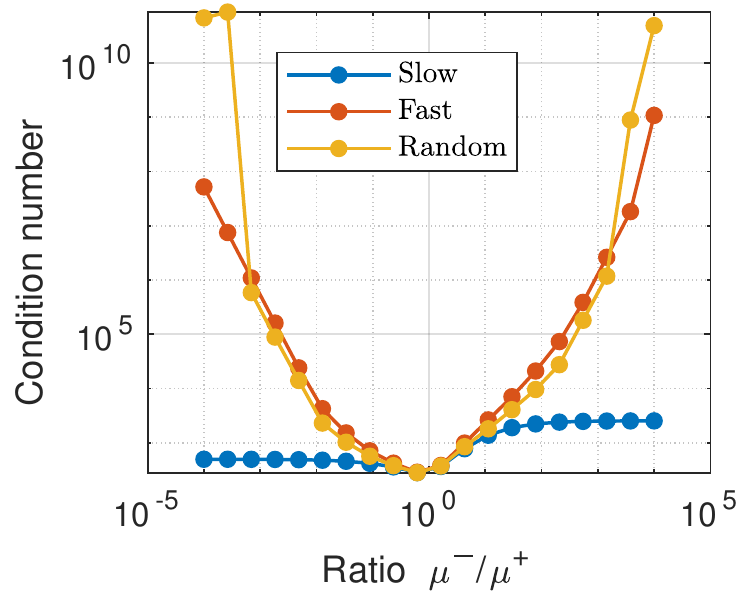}
  \end{subfigure}
  \caption{\texttt{conditioning test} in three spatial dimensions (each data point represents maximum value among $5 \times 5 \times 5 = 125$ different relative placements of an immersed interface and the computational grid).}
  \label{fig::3d::conditioning}
\end{figure}

Consider a spherical shell\footnote{As in the two-dimensional case, Dirichlet boundary conditions are enforced on the boundaries of the enclosing region} with inner and outer radii $r_i = 0.151$ and $r_e = 0.911$, and an immersed star-shaped interface described by the level-set function: 
\begin{linenomath*}
\begin{align*}
  \phi(x, y, z) = \sqrt{x^2 + y^2 + z^2} - r_0 \left(1 + \left( \frac{x^2+y^2}{x^2+y^2+z^2} \right)^2 \sum_{k=1}^{3} \beta_k \cos \left(n_k \left(\arctan \left( \frac{y}{x} \right) - \theta_k\right)\right) \right),
\end{align*} 
\end{linenomath*}
with the same parameters \eqref{eq::level-set-params} as for the two-dimensional case. The problem geometry is illustrated in Fig. \ref{fig::3d::geometry}. The exact solutions are taken to be $u^- = \sin(2x) \cos(2y) \exp(z)$ and $u^+ = \left( 16 \left( \frac{y-x}{3} \right)^5 - 20 \left( \frac{y-x}{3} \right)^3 + 5 \left( \frac{y-x}{3} \right) \right) \log \left( x+y+3 \right) \cos(z)$. In the \texttt{convergence test}, the diffusion coefficients are set to $\mu^- = 10 \left( 1 + \frac{1}{5} \cos( 2 \pi (x+y) ) \sin( 2 \pi (x-y) ) \cos(z) \right)$ and $\mu^+ = 1$. In the \texttt{conditioning test}, the grid resolution is fixed at $2^{-4}$, $\mu^+ = 1$ and $\mu^-$ is varied from $10^{-4}$ to $10^{4}$. The test results are presented in Fig. \ref{fig::3d::accuracy} and \ref{fig::3d::conditioning} ,where each data point is obtained as the maximum (worse) value among $5\times 5 \times 5 = 125$ different relative placements of the immersed interface on the computational grid. As for the two dimensional case, section \ref{sec::Analysis} will draw some conclusions from these results.

\subsection{Analysis} \label{sec::Analysis}
From the results presented in sections \ref{sec::2dTests} and \ref{sec::3dTests}, it is clear that the numerical schemes have the same behavior in two and three spatial dimensions. The \texttt{convergence test} results (see Fig. \ref{fig::2d::accuracy} and \ref{fig::3d::accuracy}) indicate that, for a moderate diffusion coefficient ratio, all three schemes have comparable convergence properties: the numerical solutions are second-order accurate with first-order accurate gradients in the $L^\infty$-norm. The condition number scales with the grid resolution as $h^{-2}$, which is similar to the scaling of the condition number for the standard five-point (nine-point) stencil in 2D (3D). The only difference between the three schemes is the magnitude of the errors and the magnitude of the condition numbers, with the scheme \texttt{Bias Slow} giving the best results.

On the other hand, the \texttt{conditioning test} in two and three spatial dimensions demonstrate that the three schemes behaviors are drastically different when the ratio $\frac{\mu^-}{\mu^+}$ varies (Fig. \ref{fig::2d::conditioning} and \ref{fig::3d::conditioning}). In particular, the condition numbers for the schemes \texttt{Random} and \texttt{Bias Fast} grow unboundedly as the ratio of the diffusion coefficients either decreases or increases away from 1. As a result, the magnitude of the errors in the solution and its gradient grow significantly. We also note that, for approximately $\frac{\mu^-}{\mu^+} > 10$ and $\frac{\mu^-}{\mu^+} < 10^{-1}$, the linear solver is not able to invert the resulting linear system in a given number of iterations (we set that number to 50 in those numerical examples). In contrast, the condition number for the scheme \texttt{Bias Slow} converges to finite values as $\frac{\mu^-}{\mu^+} \rightarrow 0$ or $\frac{\mu^-}{\mu^+} \rightarrow \infty$. As a result, the linear solver is able to invert the resulting linear sytem for any values of $\frac{\mu^-}{\mu^+}$ (the number of iterations depends only on the grid resolution). Moreover, the errors of the numerical solutions and their gradients are only moderately affected by small or large ratios $\frac{\mu^-}{\mu^+}$.

\reviewerTwo{
\subsection{Application to adaptive quadtree and octree grids}\label{sec::example:adaptive}
Thanks to its fairly small stencil, it is simple to apply the proposed scheme in the context of adaptive grids. In this section, we demonstrate perhaps the easiest way of doing so: we consider adaptive Cartesian quadtree (octree) grids that are locally uniform around immersed interfaces. In the regions where an adaptive grid is non-uniform (and which are away from immersed interfaces), we use the second-order accurate superconvergent finite difference scheme of \cite{Min;Gibou:07:A-second-order-accur}, while in the regions close to immersed interfaces (and where the grid is locally uniform), we use the proposed scheme for imposing interface jump conditions. Note that the discretization of \cite{Min;Gibou:07:A-second-order-accur} and the one described in this work reduce to the standard 5-point (9-point in three spatial dimensions) stencil on uniform grids and in the absence of immersed interfaces. This fact makes the combination of the two discretizations seamless.}

\reviewerTwo{
In this example, we consider 10 clusters of small star-shaped uniformly charged dielectric particles in a $2$m$\times$2m vacuum domain and compute the electric field generated by this configuration. The number of particles in each cluster varies from 3 to 10, the average size of clusters is 1 cm, particles have between 2 and 6 bumps and sizes in the range $[0.1; 1]$ mm. The exact problem geometry description is provided in \ref{sec::adaptivegridgeometry}. The absolute permittivity of particles is $10\varepsilon_0$ and their charge densities are $\pm10^5 \varepsilon_0$ (the sign is assigned to each particle randomly), where $\varepsilon_0$ is the vacuum permittivity. The domain boundaries are assumed to be a good conductor. Thus, the electric potential $\varphi \equiv u$ satisfies the boundary value problem \eqref{eq::poisson}-\eqref{eq::jump_flux} with parameters $k^\pm = \alpha = \beta = 0$, $\mu^- = 10$, $\mu^+ = 1$, $f^- = \pm 10^5$ (depending which particle is considered), $f^+ = 0$, $g =0$, where we denote the particles as $\Omega^-$ and the vacuum as $\Omega^+$.}

\begin{figure}[!h]
  \begin{subfigure}[t]{0.49\textwidth}
      \caption{}
    \includegraphics[width=0.95\textwidth]{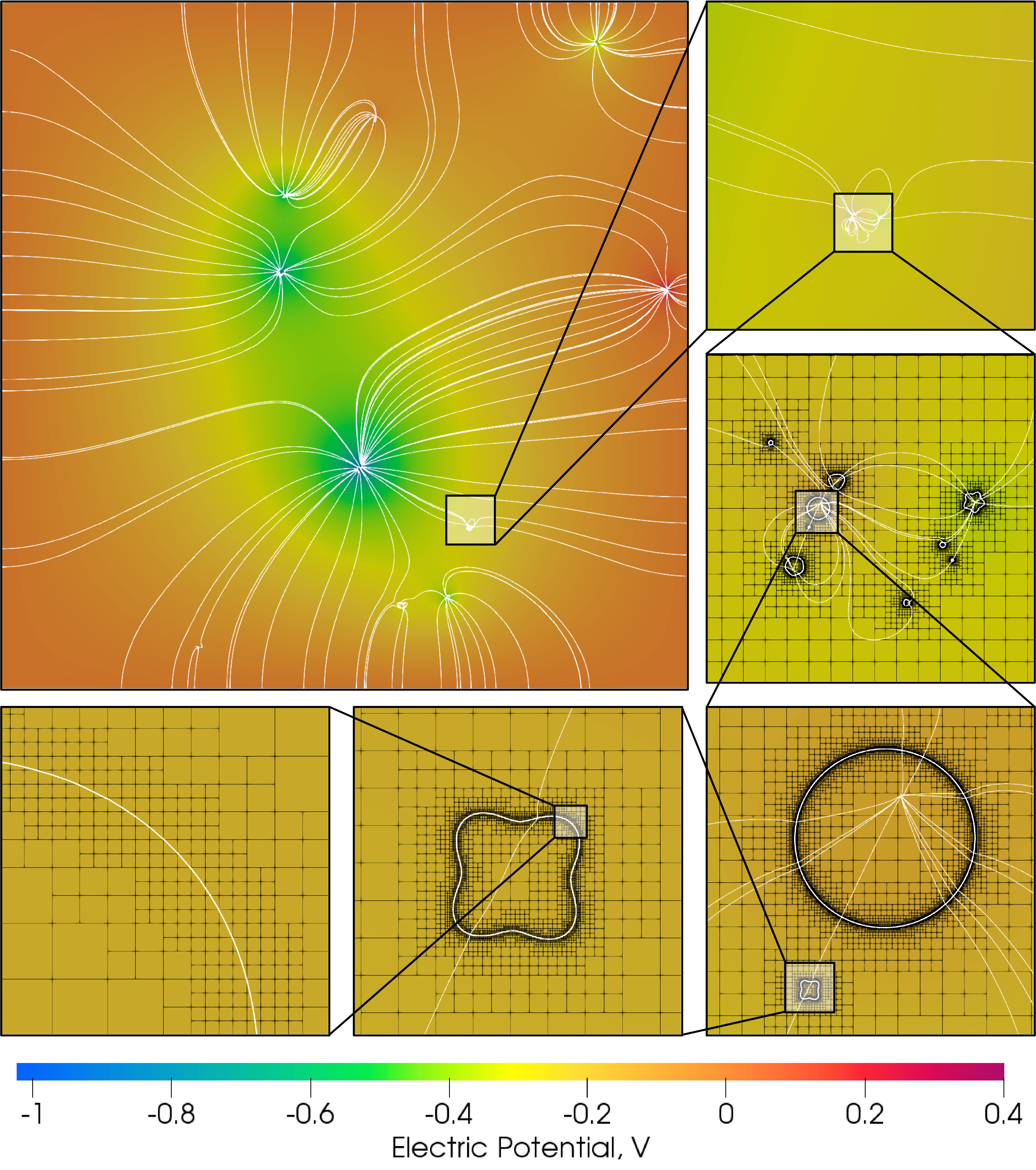}
  \end{subfigure}
  \begin{subfigure}[t]{0.49\textwidth}
      \caption{}
    \includegraphics[width=0.95\textwidth]{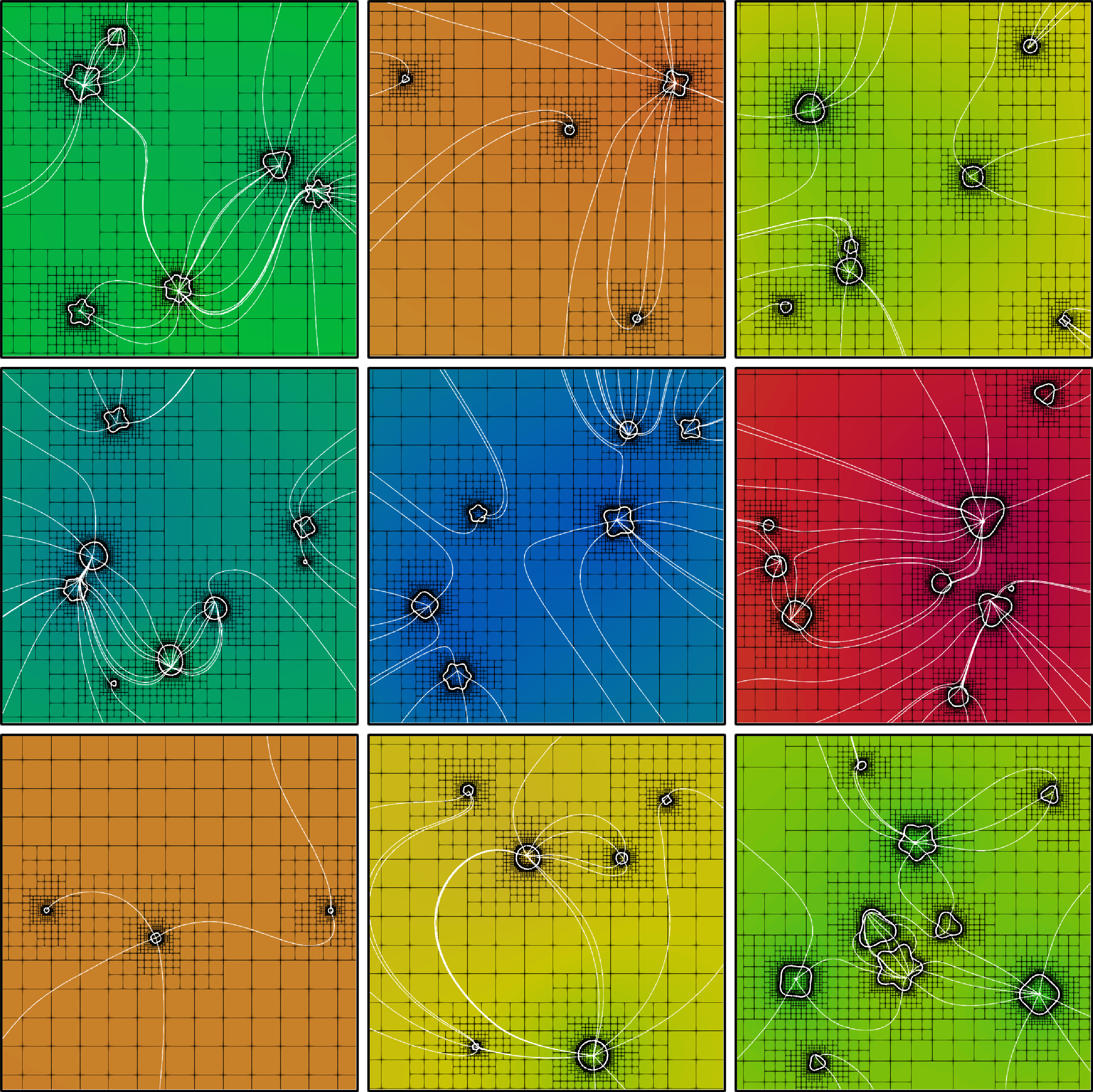}
  \end{subfigure}
  \caption{Application of the present method in the context of adaptive grids for computing electric field around clusters of charged dielectric particles: (a) The entire computational domain with successive zoom-ins for one of the clusters; (b) Zoom-ins for all other clusters. The color map indicates the electric potential and thin lines represent the electric field lines. The thicker white lines represent particles.}
  \label{fig::2d::adaptive}
\end{figure}

\reviewerTwo{
Figure \ref{fig::2d::adaptive} depicts the electric potential $\phi$ and the electric field $\vect{E} = -\nabla \phi$ computed on an adaptive grid with the coarsest and finest mesh sizes corresponding to uniform resolutions of $2^{10} \times 2^{10}$ and $2^{20} \times 2^{20}$ grid points, respectively. Such a computational grid contains $2754021$ points, which is approximately just $0.00025\%$ of the total number of points in a uniform $2^{20} \times 2^{20}$ grid.}

\vspace{.25cm}
\mylinelabel{rev2:fullyadaptive}\reviewerTwo{
\noindent\textbf{Remark}:
While the strategy of enforcing adaptive grids to be locally uniform around immersed interfaces is quite common and justified from the point of view of accuracy in many situations, it may not be very efficient for certain applications; in particular those where parts of the interface has high curvature while other parts are rather flat. However, it seems straightforward to adapt the described method to fully-adaptive (non-uniform along immersed interfaces) and Voronoi grids (including Voronoi partitions generated based on adaptive quadtree and octree grids). This will be considered in future works.}

\section{Conclusions}
We have presented a simple finite volume numerical method for solving Elliptic equations with jump conditions across irregular interfaces that are implicitly represented by a level-set function on Cartesian grids. Second-order accurate solutions and first-order accurate gradients are obtained in the $L^\infty$-norm.  The linear system is non-symmetric but the condition number is bounded, regardless of the ratio of the diffusion coefficients, so that the linear system can be inverted in a constant number of iterations that depends only on the grid resolution: the condition number scales as $\Oof{h^{-2}}$, similarly to the linear system obtained from the standard five-point stencil.

\mylinelabel{rev1:1} \reviewerOne{Future work will be focused on the analysis of the numerical scheme and its properties}, \mylinelabel{rev2:future}\reviewerTwo{the improvement to a superconvergent scheme (i.e, with second-order accurate gradients) and the extension to fully-adaptive (non-uniform along immersed interfaces) grids.}

\section*{Acknowledgement}
This research was supported by ARO W911NF-16-1-0136 and ONR N00014-17-1-2676.

\section*{References}
\bibliographystyle{abbrv}
\addcontentsline{toc}{section}{\refname}
\bibliography{./references,./references_new,./references_Voronoi}

\begin{thebibliography}{10}

\bibitem{Adams;Chartier:04:New-geometric-immers}
L.~Adams and T.~Chartier.
\newblock New geometric immersed interface multigrid solvers.
\newblock {\em SIAM J. of Scientific Comput}, 25:1516--1533, 2004.

\bibitem{Adams;Chartier:05:A-comparison-of-alge}
L.~Adams and T.~Chartier.
\newblock A comparison of algebraic multigrid and geometric immersed interface
  multigrid methods for interface problems.
\newblock {\em SIAM J. of Scientific Comput}, 26:762--784, 2005.

\bibitem{Adams;Li:02:The-immersed-interfa}
L.~Adams and Z.~Li.
\newblock The immersed interface/multigrid methods for interface problems.
\newblock {\em SIAM J. of Scientific Comput}, 24:202.

\bibitem{Babuska:70:The-finite-element-m}
I.~Babu\u{s}ka.
\newblock The finite element method for elliptic equations with discontinuous
  coefficients.
\newblock {\em Computing}, 5:207--213, 1970.

\bibitem{Balay;Brown;Buschelman;etal:12:PETSc-Web-page}
S.~Balay, J.~Brown, K.~Buschelman, W.~D. Gropp, D.~Kaushik, M.~G. Knepley,
  L.~C. McInnes, B.~F. Smith, and H.~Zhang.
\newblock Petsc web page, 2012.

\bibitem{Bao;Hong;Teran;etal:07:Fracturing-rigid-mat}
Z.~Bao, J.-M. Hong, J.~Teran, and R.~Fedkiw.
\newblock Fracturing rigid materials.
\newblock {\em IEEE Trans. on Vis. and Comput. Graph}, 13:370--378, 2007.

\bibitem{Belytschko;Moes;Usui;etal:01:Arbitrary-discontinu}
T.~Belytschko, N.~Mo\"{e}s, S.~Usui, and C.~Parimi.
\newblock Arbitrary discontinuities in finite elements.
\newblock {\em Int. J. for Num. Meth. Eng}, 50:993--1013, 2001.

\bibitem{Berthelsen:04:A-decomposed-immerse}
P.~A. Berthelsen.
\newblock A decomposed immersed interface method for variable coefficient
  elliptic equations with non-smooth and discontinuous solutions.
\newblock {\em Journal of Computational Physics}, 197:364--386, 2004.

\bibitem{Bochkov;Gibou:19:A-Sharp-Computationa}
D.~Bochkov and F.~Gibou.
\newblock A sharp computational method for the simulation of the solidification
  of multicomponent alloys.
\newblock {\em In Preparation}, 2019.

\bibitem{Bochkov;Gibou:19:Solving-the-Poisson-}
D.~Bochkov and F.~Gibou.
\newblock Solving poisson-type equations with robin boundary conditions on
  piecewise smooth interfaces.
\newblock {\em J. Comput. Phys.}, 376:1156--1198, 2019.

\bibitem{Brennen:2005aa}
C.~E. Brennen.
\newblock {\em Fundamentals of Multiphase Flows}, volume ISBN 0521 848040.
\newblock Cambridge University Press, 2005.

\bibitem{Chen;Min;Gibou:07:A-supra-convergent-f}
H.~Chen, C.~Min, and F.~Gibou.
\newblock A supra-convergent finite difference scheme for the {P}oisson and
  heat equations on irregular domains and non-graded adaptive {C}artesian
  grids.
\newblock {\em J. Sci. Comput.}, 31:19--60, 2007.

\bibitem{Chen;Strain:08:Piecewise-polynomial}
T.~Chen and J.~Strain.
\newblock Piecewise-polynomial discretization and krylov-accelerated multigrid
  for elliptic interface problems.
\newblock {\em Journal of Computational Physics}, 227(16):7503--7542, 2008.

\bibitem{Cisternino;Weynans:12:A-parallel-second-or}
M.~Cisternino and L.~Weynans.
\newblock A parallel second order cartesian method for elliptic interface
  problems.
\newblock {\em Commun. Comput. Phys.}, 12:1562--1587, 2012.

\bibitem{Coco;Russo:12:Second-Order-Multigr}
A.~Coco and G.~Russo.
\newblock Second order multigrid methods for elliptic problems with
  discontinuous coefficients on an arbitrary interface, i: One dimensional
  problems.
\newblock {\em Numerical Mathematics: Theory, Methods \& Applications}, 5:19,
  2012.

\bibitem{Crockett;Colella;Graves:11:A-Cartesian-grid-emb}
R.~Crockett, P.~Colella, and D.~Graves.
\newblock A cartesian grid embedded boundary method for solving the poisson and
  heat equations with discontinuous coefficients in three dimensions.
\newblock {\em J. Comput. Phys.}, 230(7):2451 -- 2469, 2011.

\bibitem{Daux;Moes;Dolbow;etal:00:Arbitrary-branched-a}
C.~Daux, N.~Mo\"{e}s, J.~Dolbow, N.~Sukumar, and T.~Belytschko.
\newblock Arbitrary branched and intersecting cracks with the extended finite
  element method.
\newblock {\em Int. J. for Num. Meth. Eng}, 48:1741--1760, 2000.

\bibitem{Egan:2018aa}
R.~Egan and F.~Gibou.
\newblock Fast and scalable algorithms for constructing solvent-excluded
  surfaces of large biomolecules.
\newblock {\em Journal of Computational Physics}, 374:91 -- 120, 2018.

\bibitem{Falgout;Yang:02:Hypre:-A-library-of-}
R.~D. Falgout and U.~M. Yang.
\newblock {\em {Hypre}: A library of high performance preconditioners}, volume
  2331.
\newblock Springer Berlin Heidelberg, 2002.

\bibitem{Fedkiw1999}
R.~Fedkiw, T.~Aslam, B.~Merriman, and S.~Osher.
\newblock A non-oscillatory eulerian approach to interfaces in multimaterial
  flows (the ghost fluid method).
\newblock {\em J. Comput. Phys.}, 152:457--492, 1999.

\bibitem{Fries;Belytschko:06:The-intrinsic-XFEM:-}
T.~Fries and T.~Belytschko.
\newblock The intrinsic {XFEM}: a method for arbitrary discontinuities without
  additional unknowns.
\newblock {\em Int. J. for Num. Meth. in Eng}, 68:1358--1385, 2006.

\bibitem{Gallinato;Poignard:15:Superconvergent-Cart}
O.~Gallinato and C.~Poignard.
\newblock Superconvergent {C}artesian methods for {P}oisson type equations in
  2d -- domains.
\newblock Technical report, 2015.

\bibitem{Gallinato:2017aa}
O.~Gallinato and C.~Poignard.
\newblock Superconvergent second order cartesian method for solving free
  boundary problem for invadopodia formation.
\newblock {\em Journal of Computational Physics}, 339:412 -- 431, 2017.

\bibitem{Gallinato;Poignard:17:Superconvergent-seco}
O.~Gallinato and C.~Poignard.
\newblock Superconvergent second order {C}artesian method for solving free
  boundary problem for invadopodia formation.
\newblock {\em J. Comp. Phys.}, 339:412--431, 2017.

\bibitem{Gibou;Fedkiw:05:A-fourth-order-accur}
F.~Gibou and R.~Fedkiw.
\newblock A fourth order accurate discretization for the {L}aplace and heat
  equations on arbitrary domains, with applications to the {S}tefan problem.
\newblock {\em J. Comput. Phys.}, 202(2):577 -- 601, 2005.

\bibitem{Gibou;Fedkiw;Cheng;etal:02:A-second-order-accur}
F.~Gibou, R.~Fedkiw, L.-T. Cheng, and M.~Kang.
\newblock A second-order accurate symmetric discretization of the {P}oisson
  equation on irregular domains.
\newblock {\em J. Comput. Phys.}, 176:205--227, 2002.

\bibitem{Gibou:2018aa}
F.~Gibou, R.~Fedkiw, and S.~Osher.
\newblock A review of level-set methods and some recent applications.
\newblock {\em Journal of Computational Physics}, 353:82 -- 109, 2018.

\bibitem{Gibou;Min;Fedkiw:13:High-Resolution-Shar}
F.~Gibou, C.~Min, and R.~Fedkiw.
\newblock High resolution sharp computational methods for elliptic and
  parabolic problems in complex geometries.
\newblock {\em J. Sci. Comput.}, 54:369--413, 2013.

\bibitem{Groi;Reusken:07:An-extended-pressure}
S.~Gro\'{i} and A.~Reusken.
\newblock An extended pressure finite element space for two-phase
  incompressible flows with surface tension.
\newblock {\em J. Comp. Phys}, 224:40--58, 2007.

\bibitem{Guittet;Lepilliez;Tanguy;etal:15:Solving-elliptic-pro}
A.~Guittet, M.~Lepilliez, S.~Tanguy, and F.~Gibou.
\newblock Solving elliptic problems with discontinuities on irregular domains
  -- the {V}oronoi interface method.
\newblock {\em J. Comput. Phys.}, 298:747 -- 765, 2015.

\bibitem{Guittet;Poignard;Gibou:17:A-Voronoi-Interface-}
A.~Guittet, C.~Poignard, and F.~Gibou.
\newblock A voronoi interface approach to cell aggregate
  electropermeabilization.
\newblock {\em J. Comput. Phys.}, 332:143 -- 159, 2017.

\bibitem{Guyomarch;Lee;Jeon:09:A-discontinuous-Gale}
G.~Guyomarc'h, C.-O. Lee, and K.~Jeon.
\newblock A discontinuous {G}alerkin method for elliptic interface problems
  with application to electroporation.
\newblock {\em Comm. Numer. Methods Engrg.}, 25(10):991--1008, 2009.

\bibitem{Hu;Lai;Young:15:A-hybrid-immersed-bo}
W.-F. Hu, M.-C. Lai, and Y.-N. Young.
\newblock A hybrid immersed boundary and immersed interface method for
  electrohydrodynamic simulations.
\newblock {\em J. Comput. Phys.}, 282:47--61, 2015.

\bibitem{Ji;Dolbow:04:On-strategies-for-en}
H.~Ji and J.~Dolbow.
\newblock On strategies for enforcing interfacial constraints and evaluating
  jump conditions with extended finite element method.
\newblock {\em Int. J. for Num. Meth. in Eng}, 61:204.

\bibitem{Johansen;Colella:98:A-Cartesian-Grid-Emb}
H.~Johansen and P.~Colella.
\newblock A {C}artesian grid embedded boundary method for {P}oisson equation on
  irregular domains.
\newblock {\em J. Comput. Phys.}, 147:60--85, 1998.

\bibitem{Jr.;Wang;Sifakis;etal:12:A-second-order-virtu}
J.~L.~H. Jr., L.~Wang, E.~Sifakis, and J.~M. Teran.
\newblock A second order virtual node method for elliptic problems with
  interfaces and irregular domains in three dimensions.
\newblock {\em J. Comput. Phys.}, 231(4):2015 -- 2048, 2012.

\bibitem{Kurz;Fisher:98:Fundamentals-of-Soli}
W.~Kurz and D.~J. Fisher.
\newblock {\em Fundamentals of Solidification}.
\newblock Trans Tech Publication, 1998.

\bibitem{Lew;Buscaglia:08:A-discontinuous-Gale}
A.~J. Lew and G.~C. Buscaglia.
\newblock A discontinuous-{G}alerkin-based immersed boundary method.
\newblock {\em Int. J. for Num. Meth. in Eng}, 76:427--454, 2008.

\bibitem{Li:98:A-Fast-Iterative-Alg}
Z.~Li.
\newblock A fast iterative algorithm for elliptic interface problems.
\newblock {\em SIAM J. Numer. Anal.}, 35:230--254, 1998.

\bibitem{Li;Ito:06:The-Immersed-Interfa}
Z.~Li and K.~Ito.
\newblock {\em The Immersed Interface Method -- Numerical Solutions of PDEs
  Involving Interfaces and Irregular Domains}, volume~33.
\newblock SIAM Frontiers in Applied mathematics, 2006.

\bibitem{Liu;Fedkiw;Kang:00:A-boundary-capturing}
X.-D. Liu, R.~P. Fedkiw, and M.~Kang.
\newblock A boundary capturing method for {P}oisson's equation on irregular
  domains.
\newblock {\em J. Comput. Phys.}, 160:151--178, 2000.

\bibitem{Min;Gibou:07:Geometric-integratio}
C.~Min and F.~Gibou.
\newblock {Geometric integration over irregular domains with application to
  level-set methods}.
\newblock {\em J. Comput. Phys.}, 226:1432--1443, 2007.

\bibitem{Min;Gibou:07:A-second-order-accur}
C.~Min and F.~Gibou.
\newblock A second order accurate level set method on non-graded adaptive
  {C}artesian grids.
\newblock {\em J. Comput. Phys.}, 225(1):300--321, 2007.

\bibitem{Mirzadeh;Theillard;Gibou:11:A-second-order-discr}
M.~Mirzadeh, M.~Theillard, and F.~Gibou.
\newblock A second-order discretization of the nonlinear {P}oisson--{B}oltzmann
  equation over irregular geometries using non-graded adaptive {C}artesian
  grids.
\newblock {\em J. Comput. Phys.}, 230(5):2125 -- 2140, 2011.

\bibitem{Mirzadeh;Theillard;Helgadottir;etal:12:An-Adaptive-Finite-D}
M.~Mirzadeh, M.~Theillard, A.~Helgadottir, D.~Boy, and F.~Gibou.
\newblock An adaptive, finite difference solver for the nonlinear
  {P}oisson-{B}oltzmann equation with applications to biomolecular
  computations.
\newblock {\em Communications in Computational Physics}, 13(1):150--173, 2012.

\bibitem{Mistani:2019aa}
P.~Mistani, A.~Guittet, C.~Poignard, and F.~Gibou.
\newblock A parallel voronoi-based approach for mesoscale simulations of cell
  aggregate electropermeabilization.
\newblock {\em Journal of Computational Physics}, 380:48 -- 64, 2019.

\bibitem{Moes;Cloirec;Cartraud;etal:03:A-computational-appr}
N.~Mo\"{e}s, M.~Cloirec, P.~Cartraud, and J.~Remacle.
\newblock A computational approach to handle complex microstructure geometries.
\newblock {\em Comput. Methods Appl. Mech. Eng}, 192:3162--3177, 2003.

\bibitem{Moes;Dolbow;Belytschko:99:A-finite-element-met}
N.~Mo\"{e}s, J.~Dolbow, and T.~Belytschko.
\newblock A finite element method for crack growth without remeshing.
\newblock {\em Int. J. for Num. Meth. Eng}, 46:131--150, 1999.

\bibitem{Molino;Bao;Fedkiw:04:A-Virtual-Node-Algor}
N.~Molino, J.~Bao, and R.~Fedkiw.
\newblock A virtual node algorithm for changing mesh topology during
  simulation.
\newblock {\em ACM Trans. Graph. (SIGGRAPH Proc.)}, 23:385--392, 2004.

\bibitem{Ng;Chen;Min;etal:09:Guidelines-for-Poiss}
Y.~T. Ng, H.~Chen, C.~Min, and F.~Gibou.
\newblock Guidelines for {P}oisson solvers on irregular domains with
  {D}irichlet boundary conditions using the {G}host {F}luid {M}ethod.
\newblock {\em Journal of Scientific Computing}, 41(2):300--320, May 2009.

\bibitem{Ng;Min;Gibou:09:An-efficient-fluid--}
Y.~T. Ng, C.~Min, and F.~Gibou.
\newblock An efficient fluid--solid coupling algorithm for single-phase flows.
\newblock {\em J. Comput. Phys.}, 228(23):8807 -- 8829, 2009.

\bibitem{Oevermann;Scharfenberg;Klein:09:A-sharp-interface-fi}
M.~Oevermann, C.~Scharfenberg, and R.~Klein.
\newblock A sharp interface finite volume method for elliptic equations on
  {C}artesian grids.
\newblock {\em J. Comp. Phys}, 228:5184--5206, 2009.

\bibitem{Osher;Fedkiw:03:Level-Set-Methods-an}
S.~Osher and R.~Fedkiw.
\newblock {\em Level Set Methods and Dynamic Implicit Surfaces}.
\newblock Springer, 2003.

\bibitem{Osher;Sethian:88:Fronts-propagating-w}
S.~Osher and J.~A. Sethian.
\newblock Fronts propagating with curvature dependent speed: algorithms based
  on {H}amilton-{J}acobi formulations.
\newblock {\em J. Comput. Phys.}, 79(1):12--49, 1988.

\bibitem{Papac;Gibou;Ratsch:10:Efficient-symmetric-}
J.~Papac, F.~Gibou, and C.~Ratsch.
\newblock Efficient symmetric discretization for the {P}oisson, heat and
  {S}tefan-type problems with {R}obin boundary conditions.
\newblock {\em J. Comput. Phys.}, 229:875--889, 2010.

\bibitem{Richardson;Hegemann;Sifakis;etal:11:An-XFEM-method-for-m}
C.~Richardson, J.~Hegemann, E.~Sifakis, J.~Hellrung, and J.~Teran.
\newblock An {XFEM} method for modeling geometrically elaborate crack
  propagation in brittle materials.
\newblock {\em Int. J. for Num. Meth. in Eng}, 88:1042--1065, 2011.

\bibitem{Schwartz;Barad;Colella;etal:06:A-Cartesian-grid-emb}
P.~Schwartz, M.~Barad, P.~Colella, and T.~Ligocki.
\newblock A {C}artesian grid embedded boundary method for the heat equation and
  {P}oisson's equation in three dimensions.
\newblock {\em J. Comp. Phys.}, 211(2):531--550, 2006.

\bibitem{Sethian:96:Level-set-methods}
J.~A. Sethian.
\newblock {\em Level set methods}, volume~3 of {\em Cambridge Monographs on
  Applied and Computational Mathematics}.
\newblock Cambridge University Press, Cambridge, 1996.
\newblock Evolving interfaces in geometry, fluid mechanics, computer vision,
  and materials science.

\bibitem{Shortley;Weller:38:Numerical-solution-o}
G.~H. Shortley and R.~Weller.
\newblock Numerical solution of {L}aplace's equation.
\newblock {\em J. Appl. Phys.}, 9:334--348, 1938.

\bibitem{Sifakis;Der;Fedkiw:07:Arbitrary-cutting-of}
E.~Sifakis, K.~Der, and R.~Fedkiw.
\newblock Arbitrary cutting of deformable tetrahedralized objects.
\newblock {\em In Proceedings of SIGGRAPH}, 2007:73--80, 2007.

\bibitem{Theillard;Gibou;Pollock:14:A-Sharp-Computationa}
M.~Theillard, F.~Gibou, and T.~Pollock.
\newblock A sharp computational method for the simulation of the solidification
  of binary alloys.
\newblock {\em J. Sci. Comput.}, 63:330--354, 2014.

\bibitem{Bos;Gravemeier:09:Numerical-simulation}
F.~van~der Bos and V.~Gravemeier.
\newblock Numerical simulation of premixed combustion using an enriched finite
  element method.
\newblock {\em J. Comp. Phys}, 228:3605--3624, 2009.

\bibitem{Wiegmann;Bube:00:The-explicit-jump-im}
A.~Wiegmann and K.~Bube.
\newblock The explicit jump immersed interface method: finite difference method
  for pdes with piecewise smooth solutions.
\newblock {\em SIAM J. Numer. Anal.}, 37(3):827--862, 2000.

\bibitem{Xia:2014aa}
K.~Xia, X.~Feng, Z.~Chen, Y.~Tong, and G.-W. Wei.
\newblock Multiscale geometric modeling of macromolecules i: Cartesian
  representation.
\newblock {\em Journal of Computational Physics}, 257:912 -- 936, 2014.

\bibitem{Xu;Shi;Hu;Etal:20:A-level-set-immersed}
J.-J. Xu, W.~Shi, W.-F. Hu, and J.-J. Huang.
\newblock A level-set immersed interface method for simulating the
  electrohydrodynamics.
\newblock {\em J. Comput. Phys.}, 400:108956, 2020.

\bibitem{Xu:12:An-iterative-two-flu}
S.~Xu.
\newblock An iterative two-fluid pressure solver based on the immersed
  interface method.
\newblock {\em Communications in Computational Physics}, 12(2):528--543, 2012.

\end{thebibliography}

\newpage
\appendix

\section{Geometry description for adaptive grid example}\label{sec::adaptivegridgeometry}
\reviewerTwo{
The adaptive grid example (Section \ref{sec::example:adaptive}) deals with 67 particles each of which is described by a level-set function of the form:
\begin{linenomath*}
\begin{align*}
  \phi(r, \theta) = \sqrt{(r\cos(\theta - \theta_0) - x_c)^2 + (r\sin(\theta - \theta_0)-y_c)^2} - r_0 \left( 1 + \delta \cos(m(\theta-\theta_0)) \right).
\end{align*}
\end{linenomath*}
Parameters $x_c$, $y_c$, $r_0$, $\delta$, $m$ and $\theta_0$ for each particle as well as the sign of its charge are given in the following table:
\scriptsize
  \begin{longtable}{c|c|c|c|c|c|c|c}
    Particle number & $x_c$ & $y_c$ & $r_0$ & $\delta$ & $m$ & $\theta_0$ & Charge \\
    \hline
    1 & 0.294320 & -0.731980 & 0.000948 & 0.177428 & 5 & 3.779031 & +\\ 
    2 & 0.292603 & -0.722570 & 0.000196 & 0.120466 & 2 & 0.575194 & +\\ 
    3 & 0.296621 & -0.730048 & 0.000553 & 0.195375 & 3 & 5.983689 & +\\ 
    4 & 0.301392 & -0.723932 & 0.000374 & 0.161709 & 3 & 1.178485 & +\\ 
    5 & 0.293268 & -0.730192 & 0.000798 & 0.193548 & 3 & 6.112317 & -\\ 
    6 & 0.289541 & -0.732616 & 0.000755 & 0.088156 & 4 & 2.266495 & -\\ 
    7 & 0.290527 & -0.736414 & 0.000341 & 0.140005 & 3 & 6.140219 & -\\ 
    8 & 0.295213 & -0.726113 & 0.000792 & 0.121836 & 5 & 0.884964 & -\\ 
    9 & 0.300884 & -0.733236 & 0.000857 & 0.078646 & 4 & 1.588758 & -\\ 
    10 & 0.935243 & 0.156877 & 0.000462 & 0.026599 & 4 & 1.152778 & +\\ 
    11 & 0.936367 & 0.165333 & 0.000946 & 0.129853 & 3 & 2.585702 & +\\ 
    12 & 0.937713 & 0.161907 & 0.000114 & 0.111576 & 3 & 3.069164 & +\\ 
    13 & 0.927721 & 0.160662 & 0.000653 & 0.059185 & 4 & 4.986538 & -\\ 
    14 & 0.926407 & 0.164860 & 0.000243 & 0.009826 & 2 & 3.613796 & -\\ 
    15 & 0.939303 & 0.170991 & 0.000443 & 0.141560 & 3 & 2.988958 & +\\ 
    16 & 0.926755 & 0.162948 & 0.000482 & 0.025398 & 5 & 3.380732 & +\\ 
    17 & 0.936908 & 0.160965 & 0.000713 & 0.179751 & 3 & 0.252419 & +\\ 
    18 & 0.934454 & 0.162162 & 0.000448 & 0.011111 & 6 & 2.899990 & +\\
    19 & 0.160228 & -0.743358 & 0.000329 & 0.052804 & 6 & 4.507911 & +\\ 
    20 & 0.164293 & -0.747939 & 0.000792 & 0.015578 & 3 & 3.753404 & +\\ 
    21 & 0.173756 & -0.744021 & 0.000264 & 0.186352 & 4 & 6.101247 & +\\ 
    22 & 0.168733 & -0.761421 & 0.000999 & 0.001466 & 5 & 4.162976 & -\\ 
    23 & 0.170695 & -0.747981 & 0.000389 & 0.025428 & 4 & 0.952773 & -\\ 
    24 & 0.160737 & -0.760843 & 0.000209 & 0.199829 & 6 & 0.814897 & -\\ 
    25 & 0.725447 & 0.889409 & 0.000966 & 0.064583 & 3 & 5.719127 & -\\ 
    26 & 0.740474 & 0.893693 & 0.000453 & 0.004282 & 3 & 5.779404 & -\\ 
    27 & 0.736547 & 0.884784 & 0.000708 & 0.036598 & 4 & 0.871440 & -\\ 
    28 & 0.723810 & 0.875899 & 0.000367 & 0.044844 & 3 & 0.309913 & -\\ 
    29 & 0.742820 & 0.874970 & 0.000306 & 0.191154 & 4 & 1.616747 & -\\ 
    30 & 0.728138 & 0.878376 & 0.000857 & 0.026387 & 4 & 5.913726 & -\\ 
    31 & 0.728254 & 0.880038 & 0.000451 & 0.069948 & 6 & 2.499319 & +\\ 
    32 & 0.052907 & -0.344586 & 0.000660 & 0.184134 & 4 & 2.412480 & -\\ 
    33 & 0.034801 & -0.356601 & 0.000834 & 0.079140 & 4 & 4.794653 & -\\ 
    34 & 0.048034 & -0.350902 & 0.000961 & 0.175580 & 4 & 1.000997 & -\\ 
    35 & 0.038434 & -0.350375 & 0.000546 & 0.119999 & 5 & 5.330543 & -\\ 
    36 & 0.037001 & -0.361473 & 0.000859 & 0.104759 & 6 & 2.104655 & -\\ 
    37 & 0.048676 & -0.344655 & 0.000590 & 0.018488 & 3 & 3.979605 & -\\ 
    38 & -0.182697 & 0.433937 & 0.000642 & 0.191264 & 5 & 0.196232 & -\\ 
    39 & -0.180690 & 0.449481 & 0.000499 & 0.107538 & 4 & 3.958288 & +\\ 
    40 & -0.171622 & 0.442325 & 0.000714 & 0.162207 & 3 & 4.889946 & -\\ 
    41 & -0.177232 & 0.435290 & 0.000706 & 0.108419 & 6 & 2.765306 & +\\ 
    42 & -0.182553 & 0.446933 & 0.000927 & 0.139933 & 5 & 0.682710 & -\\ 
    43 & -0.169332 & 0.440621 & 0.000652 & 0.197255 & 6 & 0.524252 & -\\ 
    44 & 0.082096 & 0.665721 & 0.000309 & 0.018026 & 6 & 3.925875 & +\\ 
    45 & 0.070903 & 0.669204 & 0.000229 & 0.189876 & 3 & 1.973639 & +\\ 
    46 & 0.086667 & 0.652859 & 0.000256 & 0.009732 & 4 & 5.337312 & -\\ 
    47 & 0.089332 & 0.668903 & 0.000778 & 0.189929 & 4 & 0.506358 & +\\ 
    48 & -0.416502 & -0.873554 & 0.000168 & 0.038622 & 2 & 4.772787 & +\\ 
    49 & -0.435866 & -0.873537 & 0.000167 & 0.040327 & 6 & 4.357852 & +\\ 
    50 & -0.428412 & -0.875426 & 0.000329 & 0.067637 & 4 & 4.422504 & -\\ 
    51 & 0.367410 & -0.528637 & 0.000239 & 0.004636 & 6 & 1.018504 & -\\ 
    52 & 0.364188 & -0.533808 & 0.000271 & 0.072883 & 3 & 2.058938 & -\\ 
    53 & 0.357943 & -0.522965 & 0.000682 & 0.097692 & 3 & 4.725618 & +\\ 
    54 & 0.354170 & -0.530541 & 0.000809 & 0.035313 & 5 & 0.726723 & -\\ 
    55 & 0.355478 & -0.527060 & 0.000107 & 0.166186 & 4 & 5.450653 & +\\ 
    56 & 0.356304 & -0.525395 & 0.000985 & 0.004988 & 2 & 3.471888 & +\\ 
    57 & 0.368282 & -0.530008 & 0.000107 & 0.133467 & 5 & 5.475823 & -\\ 
    58 & 0.370185 & -0.524812 & 0.000830 & 0.145549 & 5 & 6.126808 & -\\ 
    59 & 0.352045 & -0.519527 & 0.000198 & 0.011507 & 4 & 3.886958 & +\\ 
    60 & -0.187080 & 0.207027 & 0.000928 & 0.121431 & 2 & 1.730282 & +\\ 
    61 & -0.177875 & 0.213721 & 0.000126 & 0.001767 & 3 & 0.076936 & +\\ 
    62 & -0.177947 & 0.216082 & 0.000647 & 0.130215 & 4 & 6.002824 & -\\ 
    63 & -0.193546 & 0.211842 & 0.000749 & 0.088286 & 6 & 4.366932 & -\\ 
    64 & -0.190933 & 0.205428 & 0.000170 & 0.122096 & 3 & 0.951230 & -\\ 
    65 & -0.184008 & 0.210547 & 0.000752 & 0.030592 & 3 & 3.986605 & -\\ 
    66 & -0.192313 & 0.214133 & 0.000925 & 0.029170 & 4 & 1.284015 & -\\ 
    67 & -0.190761 & 0.223399 & 0.000728 & 0.195173 & 4 & 5.881409 & -\\ 
  \end{longtable}
}

\end{document}